\documentclass[twocolumn,showpacs,
aps,superscriptaddress,
prd,notitlepage,showkeys,
nofootinbib]{revtex4-1}

\usepackage{amssymb}
\usepackage{amsmath}
\usepackage{graphicx}
\usepackage{dcolumn}
\usepackage[colorlinks,urlcolor=blue,citecolor=blue,linkcolor=blue]{hyperref}
\usepackage{color,units}
\usepackage[dvipsnames]{xcolor} 
\usepackage{lineno}
\usepackage{xspace}
\usepackage{longtable} 
\usepackage{float} 

\usepackage{amsfonts,wasysym,epsfig,verbatim,subfigure,bm,mathrsfs,lipsum}
\begin{document}
\newcommand{\IUCAA}{Inter-University Centre for Astronomy and
Astrophysics, Post Bag 4, Ganeshkhind, Pune 411 007, India}

\newcommand{\NIKHEF}{Nikhef - National Institute for Subatomic Physics, Science Park, 1098 XG Amsterdam, Netherlands}

\newcommand{\IGSPUU}{Institute for Gravitational and Subatomic Physics, Utrecht University, Princetonplein 1, 3584 CC Utrecht, The Netherlands}

\newcommand{\UVA}{Institute for High-Energy Physics, University of Amsterdam, Science Park, 1098 XG Amsterdam, Netherlands}

\newcommand{\MPI}{Max-Planck-Institut f{\"u}r Gravitationsphysik (Albert-Einstein-Institut), D-30167 Hannover, Germany}

\newcommand{\LBNZ}{Leibniz Universit{\"a}t Hannover, D-30167 Hannover, Germany}

\newcommand{\CHECK}[1]{{\color{red}~\textsf{#1}}}
\title{Imprint of black hole area quantization and Hawking radiation on inspiraling binary}
\author{Sayak Datta}\email{skdatta@iucaa.in} 
\affiliation{\IUCAA}\affiliation{\MPI}\affiliation{\LBNZ}
\author{Khun Sang Phukon}\email{k.s.phukon@nikhef.nl}
\affiliation{\NIKHEF}
\affiliation{\UVA}
\affiliation{\IGSPUU}
\date{\today}

\begin{abstract}

We study the potential of gravitational wave astronomy to observe the quantum aspects of black holes. According to Bekenstein’s quantization, we find that black hole area discretization can have observable imprints on the gravitational wave signal from an inspiraling binary black hole. We study the impact of quantization on tidal heating. We model the absorption lines and compute gravitational wave flux due to tidal heating in such a case. By including the quantization we find the dephasing of the gravitational wave, to our knowledge it has never been done before. We discuss the observability of the phenomena in different parameter ranges of the binary. We show that in the inspiral, it leads to vanishing tidal heating for the high spin values. Therefore measuring non-zero tidal heating can rule out area quantization. We also argue that if area quantization is present in nature then our current modeling with reflectivity can possibly probe the Hawking radiation which may bring important information regarding the quantum nature of gravity.

\end{abstract}

\maketitle

\section{Introduction}

Since the observation of GW190514 \cite{LIGOScientific:2018mvr} gravitational
wave (GW) astronomy has grown to be one of the dominant contributors to astrophysical observations. These observations have led to an unprecedented probe of strong gravity~\cite{LIGOScientific:2019fpa}. Properties of vacuum spacetime, propagation of GW, violation of Lorentz invariance have been tested rigorously, which has resulted in stringent bounds on the mass of the graviton and violations of Lorentz invariance \cite{TheLIGOScientific:2016pea, TheLIGOScientific:2016src, Abbott:2017vtc}. We also have begun to investigate the properties of these dark compact objects, both in inspiraling binary as well as in the merger. 

To resolve the information-loss paradox, Planck scale modifications of black hole (BH) horizons and BH structure have been proposed \cite{Lunin:2001jy, Almheiri:2012rt}. Other exotic compact object (ECOs) have also been proposed in the literature \cite{Mazur:2004fk, Liebling:2012fv}. To probe the nature of the compact objects in binary, several tests have been proposed too \cite{Cardoso:2016rao, Cardoso:2016oxy,Maselli:2017cmm, Tsang:2019zra,Abedi:2016hgu, Westerweck:2017hus, Cardoso:2019rvt,Tsang:2019zra,Abedi:2016hgu, Westerweck:2017hus,Cardoso:2017cfl, Sennett:2017etc, Brustein:2020tpg, Brustein:2021bnw, Krishnendu:2017shb, Datta:2019euh, Datta:2021hvm}.

In General Relativity, the horizon of the classical BHs (CBHs) are perfect absorbers \cite{MembraneParadigm,Damour_viscous,Poisson:2009di,Cardoso:2012zn}. This is due to the causal structure of the geometry of CBH. This null surface, which is the defining feature of a CBH, is a one-way membrane. Due to the nature of the horizon, the reflectivity of the CBH horizon is considered to be zero \cite{Berti:2009kk}. But in the case of the ECOs, it is required to introduce non-zero reflectivity \cite{Cardoso:2019apo}. This, as a result, brings changes to the observable quantities. Having such scopes, we have begun to investigate the nature of these compact objects in detail. In light of this, it has also become possible to investigate the quantum gravity corrections near horizon scales observationally. 

In this work, we focus primarily on the possibility that the GWs emitted from an inspiraling binary BH merger can carry imprints regarding the quantum properties of the BHs. In some works, this question have been posed and have been addressed primarily in the context of the postmerger \cite{Cardoso:2019apo, Laghi:2020rgl, Agullo:2020hxe}. In this work, we will focus on the possibility of observing these effects in the inspiral phase.

\section{Quantum Black Holes}

It is expected that the quantum BHs (QBHs) have a discrete energy spectrum. As a result, it has been put forward that they will behave in a similar manner as the excited atoms \cite{Mukhanov:1986me, Bekenstein:1974jk, Bekenstein:1995ju, Kothawala:2008in, Davidson:2019bqu, deFreitasPacheco:2020wdg}. Bekenstein had proposed the idea based on the adiabatic invariance of BH area in classical GR and applying Bohr-Sommerfeld quantization


\begin{equation}
A_N = \alpha \ell_P^2 N,
\end{equation}
where $\ell_{P}$ is the Planck length,  N a positive integer, and $\alpha$ is a non-negative number. 

Bekenstein and Mukhanov had concluded that BHs must have a discrete spectrum of mass and investigated the emission spectrum \cite{Bekenstein:1995ju, Agullo:2010zz}. In loop quantum gravity too BH area has been shown to be quantized \cite{Rovelli:1994ge, Rovelli:1996dv, Ashtekar:1997yu, Ashtekar:2000eq, Agullo:2008yv}. Hence, the area quantization can be considered to be one of the possible signatures of quantum gravity. Interestingly, as we will demonstrate, despite the area quantization being related to the Planck scale, it can leave observable imprints on GWs.

In Ref. \cite{Agullo:2020hxe} how the area quantization leads to an absorption spectrum has been studied in detail. They have argued that the ``astrophysical BHs act as magnifying lenses,” in the sense that they bring the Planck-scale discretization of the horizon within the realm of GW observations. In \cite{Agullo:2020hxe} various mechanisms where BH area quantization may leave observable imprints in GWs has also been discussed. One such mechanism is tidal heating. In the present work, we will investigate the impact of area quantization on tidal heating in more details. Therefore, our primary focus would be the inspiral phase of a binary.

\section{Effect of area quantization}

Effect of area quantization for a spinning BH has been explicitly computed in Ref. \cite{Agullo:2020hxe, Chakraborty:2017opo}. As our work depends on it, we will shortly review the results here. A spinning BH, which is represented by a Kerr metric, is parametrized by its mass $(M)$ and angular momentum $(J)$. Since the area of the horizon $(A)$ of the BH can completely be determined in terms of $M$ and $J$, it can be used to represent $M$ in terms of $A$ and $J$:

\begin{equation}
\label{eq:classical mass}
M = \sqrt{\frac{A}{16\pi} + \frac{4\pi J^2}{A}}.
\end{equation}

Considering the area quantization and quantization of the angular momentum, $J=\hbar j$ $(0 \leq j \leq \alpha N/8\pi)$, Ref. \cite{Agullo:2020hxe} found that, 

\begin{equation}
M_{N,j} = \sqrt{\hbar}\sqrt{\frac{\alpha N}{16\pi} + \frac{4\pi j^2}{N \alpha}}.
\end{equation}
This set of $M_{N,j}$ constitutes the mass spectrum of QBH for a set $\{N,j\}$.

By focusing on the dominant GW mode $(l=2,\,m=2)$, which is most interesting for astrophysical systems (in quasicircular binary inspiral), and considering the transition $M_{N,j} \rightarrow M_{N+\Delta N,j+2}$ in Ref. \cite{Agullo:2020hxe} it is shown that the frequency at which a BH can have non-zero absorption is,

\begin{equation}
\label{eq:base equation omega}
\omega_n = n\frac{\kappa\alpha}{8\pi} + 2\Omega_H,
\end{equation}
where $\kappa$ is the surface gravity and $\Omega_H$ is the angular velocity of the horizon\footnote{When GW frequency does not match with the frequency in Eq. (\ref{eq:base equation omega}), the absorption is more likely to be suppressed rather than being exactly equal to zero. This is primarily because the particular details of the frequency dependencies of the absorption should depend on the details of the microscopic theory of quantum gravity, that one lacks).
}. Note, for higher modes of GW with $l>2$, different transitions will also contribute. This will bring out more interesting features. But throughout our current work, we will assume that Eq. (\ref{eq:base equation omega}) is valid, and we will study the consequence of it in the inspiral phase.

Assuming Hawking radiation as the decay channel, in Ref. \cite{Agullo:2020hxe} the width $(\Gamma)$ of the lines have also been calculated. We use these numerically calculated values of $\Gamma$, that has been provided to us by the authors of Ref. \cite{Agullo:2020hxe}, to find an analytical fitting function for $\Gamma$ as follows, 
\begin{equation}
\label{eq:gamma}
M\Gamma_{E} = 1.005e^{( - 6.42 + 1.8\chi^2 + 1.9\chi^{12} - .1\chi^{14})}.
\end{equation}

\begin{figure}
\centering
\includegraphics[width=\linewidth]{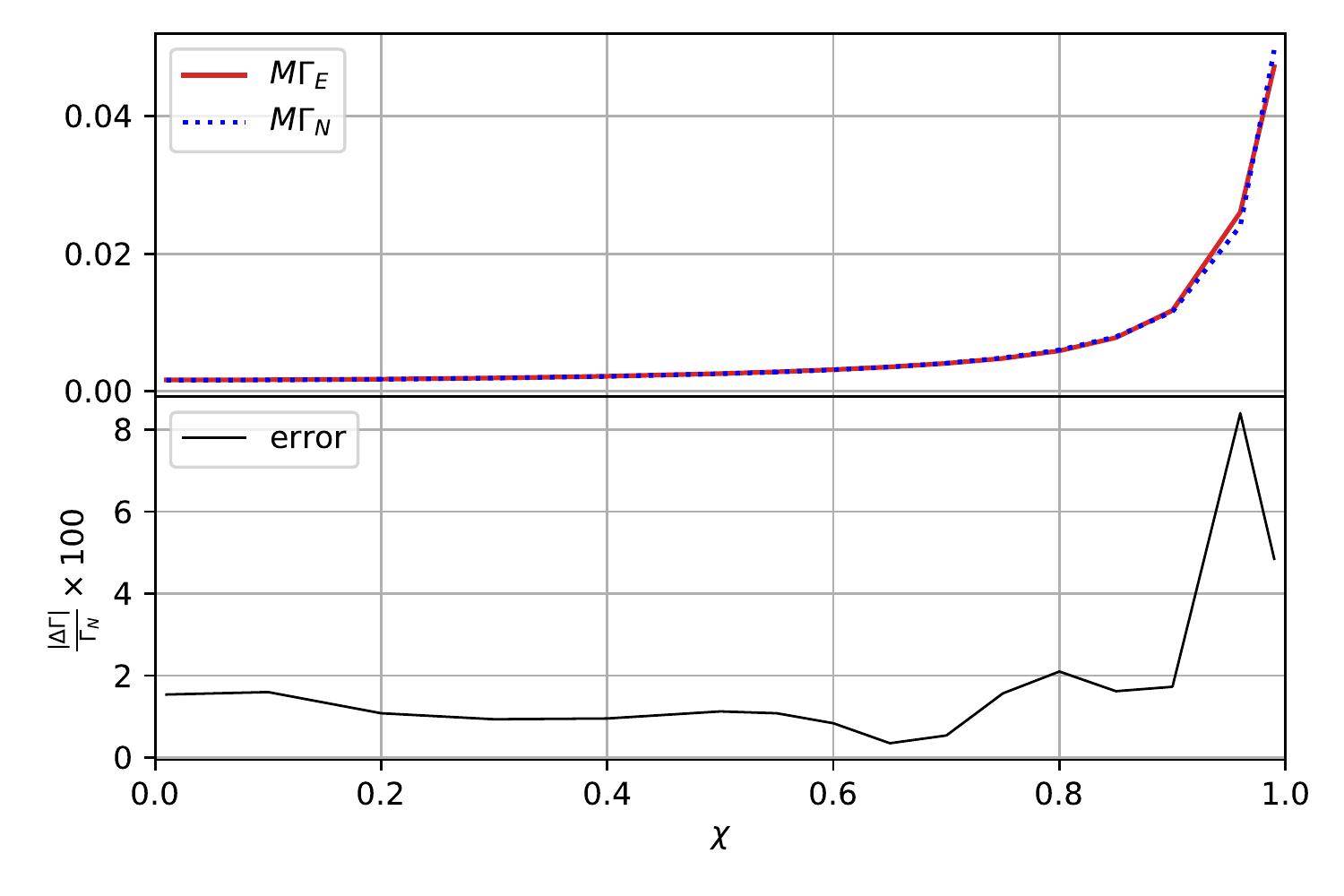}
\caption{We plot the analytical function $\Gamma_E$ (red curve) and the numerical data $\Gamma_N$ (blue dotted curve) with respect to the dimensionless spin parameter $(\chi)$ in the upper panel. In lower panel, we plot the error in the fitting. We find that for entire range of $\chi$ error is less than $8\%$, while for $\chi\leq.85$ it is less than $2\%$. Error is highest for $\chi=.9$, i.e $\sim 8\%$.\label{fig:Gamma_spin_error}}
\end{figure}

In the upper panel of the Fig. \ref{fig:Gamma_spin_error} we plot the analytical function, $\Gamma_E$, (red curve) and the numerical data, $\Gamma_N$ (blue dotted curve) with respect to the dimensionless spin parameter $(\chi)$. In the lower panel we plot the percantage fitting error defined as follows,

\begin{equation}
    |\Delta \Gamma| \equiv |\Gamma_{N} - \Gamma_{E}|,
\end{equation}
where $\Gamma_{N}$ represents the numerically evaluated values of $\Gamma$. We find that for the entire range of $\chi$, error in fitting is less than $8\%$, while for $\chi\leq.85$ it is less than $2\%$. Error is highest for $\chi=.9$, i.e $\sim 8\%$. We will use this analytic expression throughout our work.

\section{modelling the absorption: tidal heating}

A CBH absorbs radiation of frequency $\omega > m\Omega_H$ (where m is the azimuthal number of the wave
and $\Omega_H$ is the angular velocity of the horizon) but amplifies radiation of smaller frequency, due to superradiance \cite{Brito:2015oca}. When CBHs are in inspiraling binary, they expirence the other component's tidal field
during the inspiral phase. If the bodies are (at least partially) absorbing, then these tides backreact on the orbit. As a result, energy and angular momentum is transferred from their spin into the orbit. This phenomena is called tidal heating \cite{Hartle:1973zz, Hughes:2001jr, PoissonWill}. This effect is due to the dissipative nature of the CBH horizons. Modification of physics near horizon scale, as a result, can modify the TH \cite{Datta:2019epe, Datta:2020rvo}. For this reason, the TH contribution in an inspiraling binary can deviate significantly if the quantum effects near the horizon are taken into account, such as area quantization. 

As has been discussed earlier, a QBH can absorb (emit) GW only in some discrete frequency values, unlike CBH where the process is continuous across the frequency band. Therefore in the presence of area quantization tidal heating should contribute only to those frequency values that are allowed by Eq. (\ref{eq:base equation omega}). This can be modeled by assuming that the BHs have a frequency-dependent reflectivity $\mathcal{R}(f)$. For this purpose, in this section we will try to model the reflectivity of a QBH.

 The primary focus of our current work is to investigate what is the implication of such an effect in the tidal heating of a QBH in a binary, which was partially addressed in Ref. \cite{Agullo:2020hxe}. In Ref. \cite{Datta:2020rvo, Datta:2019epe} it has been established how the tidal heating in ECOs can get modified. It has been established that this can be done by defining the Horizon parameter $(H)$ and multiplying the TH flux of CBH with $H$ \cite{Maselli:2017cmm, Datta:2019euh, Sherf:2021ppp}. For horizonless bodies, $H=0$ and for CBH, $H=1$, and ``anything" in-between can be modeled assuming $0<H<1$. In Ref. \cite{Datta:2020rvo} it has been established that $H=1-|\mathcal{R}|^2$ at $\mathcal{O}(\epsilon^0)$. Where the position of the reflective surface $(r_s)$ is defined in terms of $\epsilon$ and the position of the CBH horizon $(r_H)$, as $r_s = r_H (1+\epsilon)$. In the current work, this process can be continued assuming $\epsilon = 0$ and the reflectivity $\mathcal{R}$ is frequency-dependent. Rather than constructing $\mathcal{R}(f)$, it is much easier to construct $\mathcal{H}(f)$ in the current case. We will follow that route here. Once the frequency-dependent Horizon parameter $(\mathcal{H}(f))$ is defined, frequency-dependent reflectivity can be constructed by assuming $\mathcal{H}(f)=1-|\mathcal{R}(f)|^2$. Note, in this method only $|\mathcal{R}(f)|$ can be constructed and it will not be possible to construct the phase of the reflectivity (which becomes relevant when $\epsilon \neq 0$ \cite{Chakraborty:2021gdf}). Note, once the reflectivity is constructed, irrespective of how, it can not only be used for TH computation but also for any kind of study regarding QBH. In inspiral along with TH, excited resonace modes can also contribute \cite{Maggio:2021uge, Sago:2021iku}. To study such behavior also the reflectivity constructed in this work can be used.

\begin{figure}
\centering
\includegraphics[width=\linewidth]{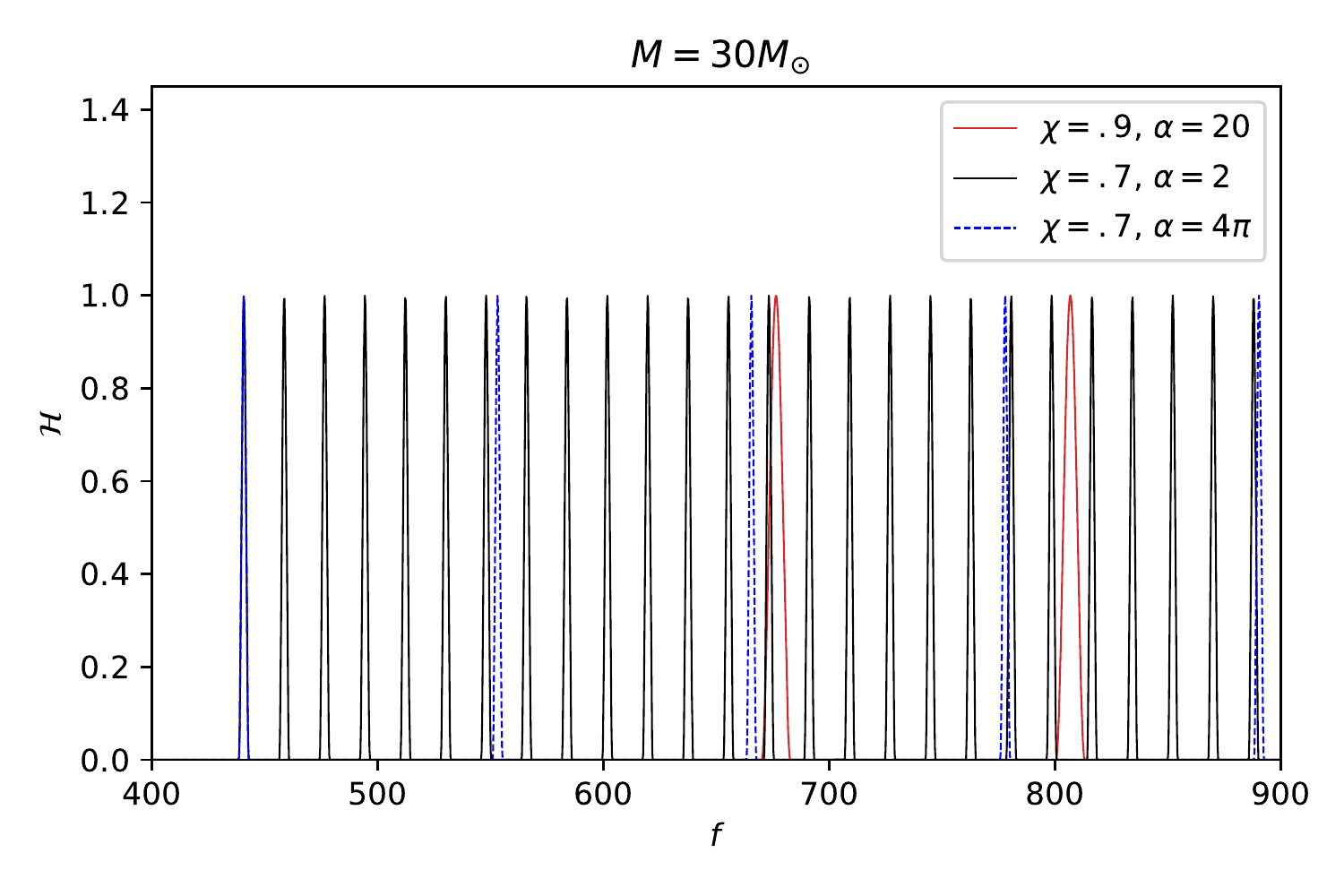}
\caption{We plot the horizon parameter $\mathcal{H}(f)$ w.r.t. the GW frequency $f$, for different spin and $\alpha$. Black and blue plot represents $\chi=.7$, where black curve has $\alpha = 2$ and blue curve represents $\alpha = 4\pi$. The first line falls on top of each other in this case, since $f_0$ depends only on the spin. With a lower value of $\alpha$, the lines in the black curves are closer to each other compared to the blue curve representing a larger $\alpha$ value. Similar phenomena happen for $\chi = .9$, represented by the red curves. Note, the width of the lines is larger for a higher spin.}
\label{fig:H}
\end{figure}

We observe that for QBH, TH will contribute only at those GW frequencies $(f)$ where $f= \omega_n/ 2\pi $. At these GW frequencies $\mathcal{H}(f_n) = 1$ should be satisfied, where $2\pi f_n \equiv \omega_n$. Since there is a non-zero line width $\Gamma$, TH should also contribute in the frequency range $f_n - \Gamma/2 < f < f_n + \Gamma/2$, following a pattern of absorption line structure. Everywhere else $\mathcal{H}(f) = 0$ should be satisfied. This can be done by constructing an absorption line profile at $f_n - \Gamma/2 < f < f_n + \Gamma/2$ for each $n$ and add them. Therefore, $\mathcal{H}(f)$ can be defined as follows:

\begin{eqnarray}
\mathcal{H}(f) =& \sum_{n=0}^{n_{max}} \mathcal{P}_n(f)\\
\mathcal{P}_n(f) =&\mathfrak{P}(f-f_n, \frac{\Gamma}{2}),
\end{eqnarray}
where $\mathfrak{P}(f-f_n, \frac{\Gamma}{2})$ is a Hann window function defined as follows,

\begin{equation}
\begin{split}
\mathfrak{P}(x, y) =\, &0,\qquad\qquad\qquad\qquad\qquad\,\,\, x\leq -y \\
&.5+.5\cos(2\pi\frac{.5x}{y}),\,\,\,\, -y\leq x\leq y\\
\,&0, \qquad\qquad\qquad\qquad\qquad\,\,\, x \geq y 
\end{split}
\end{equation}

\begin{eqnarray*}
& &\omega_n = n\frac{\kappa\alpha}{8\pi} + 2\Omega_H,\,\,n_{max} = Ceiling[\frac{8\pi}{\kappa \alpha}(\omega_{contact}-\omega_0)]\\ 
& &\kappa = \frac{\sqrt{1-\chi^2}}{2M(1+\sqrt{1-\chi^2})},\,\,\,
\Omega_H =\frac{\chi}{2M(1+\sqrt{1-\chi^2})}.
\end{eqnarray*}

In Fig. \ref{fig:H} we plot $\mathcal{H}(f)$ for multiple parameter sets. Black and blue plot represents $\chi=.7$, where black curve has $\alpha = 2$ and blue curve represents $\alpha = 4\pi$. The first line falls on top of each other in this case, since $f_0$ depends only on the spin. With a lower value of $\alpha$ in the black curve, the lines are very close to each other compared to the blue curve with larger value of $\alpha$. Similar phenomena happens for $\chi = .9$, which is represented by the red curve. Note that the width of the lines are larger for a higher spin. This can be understood by comparing the red and the black curve. This is also evident from Fig. \ref{fig:Gamma_spin_error}.

In Ref \cite{Agullo:2020hxe} critial values of $\alpha$, say $\alpha_{crit}(\chi)$, below which lines start to overlap, has been calculated as a function of spin. By taking sufficiently small $\alpha$, it is possible to get $\mathcal{H}(f) \rightarrow 1$ for the entire frequency range where $f>f_0$. Therefore, in frequency range $f>f_0$ it can mimic the result of a CBH. This happens because for $\alpha<\alpha_{crit}$ the closely spaced lines starts to overlap with each other. As a result, their contributions get added up to give a ``semi-uniform" value of $\mathcal{H}(f)$. Although for small values of $\alpha$, $H \rightarrow 1$, it can not mimic CBH result exactly. For CBH there is no lowest value of $f$ below which TH vanishes. But for QBH if $f < f_0$, then TH vanishes, even for very small values of $\alpha$. This phenomena is very different from CBH, as in CBH (unlike QBH) TH will be non-zero not only for $f>f_0$ but also for $0<f < f_0$.

\begin{figure}
\centering
\includegraphics[width=\linewidth]{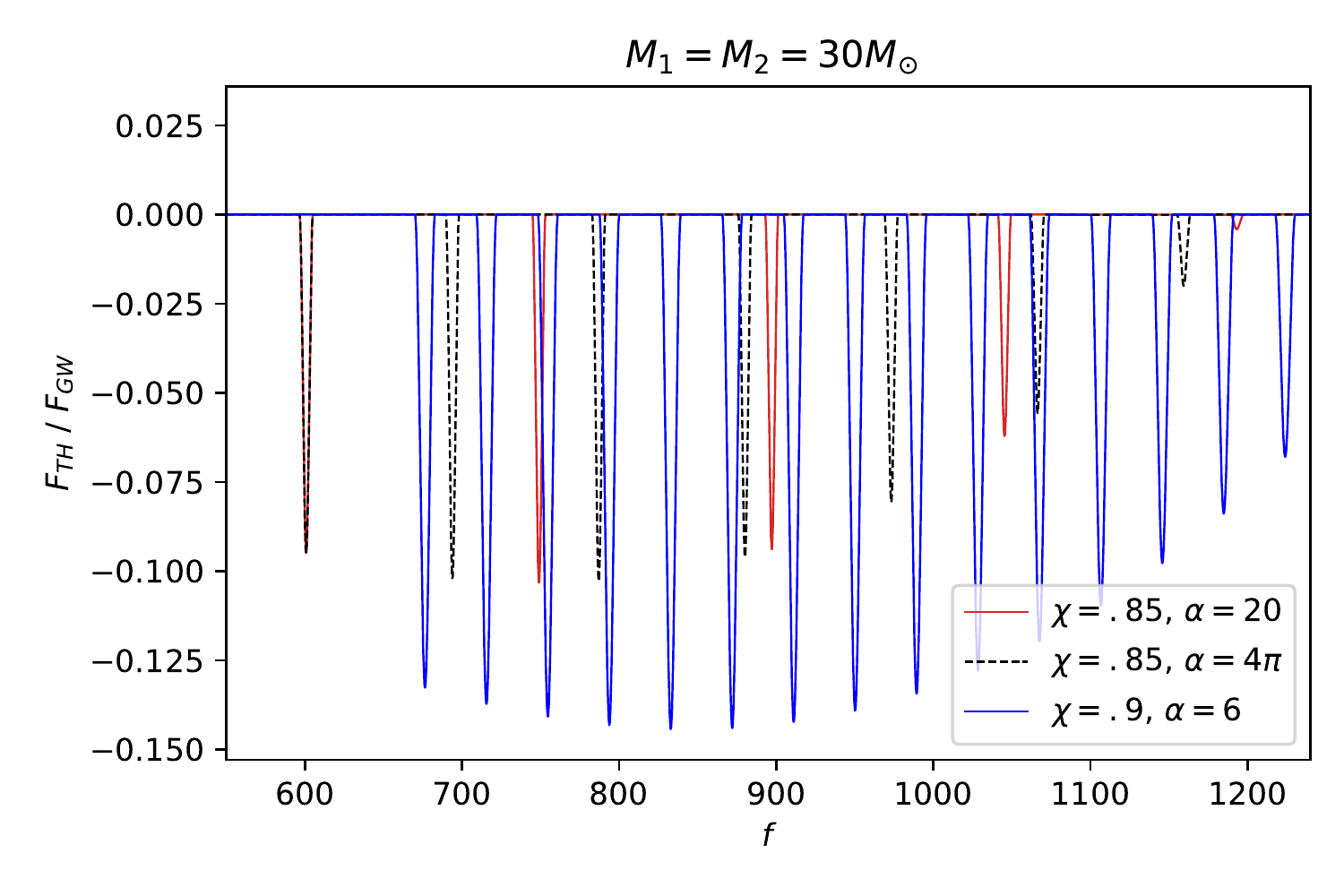}
\caption{We plot the fractional flux due to tidal heating w.r.t. the flux at infinity assuming the area quantization. As expected, the CBH flux envelopes the QBH flux. It can have both the absorption lines as well as emission lines which should be understood in the context of the superradiance of classical BH.}
\label{fig:F}
\end{figure}

Once $\mathcal{H}(f)$ is constructed, flux for TH of QBH in a binary can be defined by multiplying $\mathcal{H}(f)$ to the TH flux for CBH, namely $\mathcal{F}_{CBH}(f)$. Therefore, the flux for QBH, namely $\mathcal{F}_{QBH}(f) \equiv \mathcal{H}(f)\mathcal{F}_{CBH}(f)$. Using this definition, we express the total flux due to TH of a binary QBH in the following manner,
\begin{equation}
\mathcal{F}_{\rm TH,QBH} = \mathcal{H}^{(1)}\mathcal{F}^{(1)}_{CBH} + \mathcal{H}^{(2)} \mathcal{F}^{(2)}_{CBH},
\end{equation}
where, $M_i$ and $\chi_i$ is the mass and the dimensionless spin of the $i$th QBH, mass ratio $q=M_2/M_1$, $M=M_1+M_2$ and $\mathcal{H}^{(i)}$ is the horizon parameter, and $\mathcal{F}^{(i)}$ is flux of the $i$th body if it were a CBH (expression can be found in Ref. \cite{Alvi:2001mx, Datta:2020gem, Chatziioannou:2012gq, Chatziioannou:2016kem}).

In Fig. \ref{fig:F} we have plotted an example of fractional flux due to TH of QBH with respect to the flux at infinity. Due to the area quantization, TH flux has a discretized structure that is determined by the area quantization parameter $\alpha$. But as expected, the discretized structure follows the CBH envelope. For all of the curves the components of the binaries are equally massive,i.e. $q=1$. Individual mass of the components are $30M_{\odot}$. The red and the black curve represents $\chi=.85$ and $\alpha=20,\,4\pi$ respectively. The blue curve is for $\chi=.9$ and $\alpha=6$. Note, this plot is for representation purpose only, as we will discuss later, there will be no TH for these spin values in equal mass binary.

We deduce the GW phase involving tidal heating by using Eq.~(2.7) of Ref.~\cite{Tichy:1999pv} (see \cite{Isoyama:2017tbp} for the details). We find the phase shift due to the associated horizon absorption to be,

\begin{equation}
\begin{aligned}
\label{eq:phase correction}
\Psi^{}_{\rm THQBH} = &{} \frac{3}{128\nu} \left(\frac{1}{v}\right)^5 \left[- \frac{10 }{9 }v^5 \Psi_{5} \left(3 \log \left(v\right)+1\right) \right. \\
&- \frac{5}{168} v^7 \Psi_{5} \left(952 \nu +995\right) \\ 
&\left.+ \frac{5}{9}v^8 \left(3 \log \left(v\right)-1\right)(-4 \Psi_{8}+ \Psi_5 \psi^{}_{\text{SO}} )\right]\,,
\end{aligned}
\end{equation}
where $\Psi_5 = (\mathcal{H}^{(1)}A^{(1)}_{5}+\mathcal{H}^{(2)}A^{(2)}_{5})$ and $\Psi_8 = (\mathcal{H}^{(1)}A^{(1)}_{8}+\mathcal{H}^{(2)}A^{(2)}_{8})$, $v$ is the post-Newtonian velocity parameter, $\nu$ is the symmetric mass ratio, $\hat{L}$ is the direction of the orbital angular momentum, $\hat{S}_i$ is the direction of the $i-$th component's spin,
\begin{equation}
\begin{split}
\psi^{}_{\text{SO}} \equiv~ &\frac{1}{6} \big[\big(-56 \nu -73 \sqrt{1-4 \nu }+73\big) \big(\hat{L}.\hat{S}^{}_1\big) \chi _1 \\
&+\big(-56 \nu + 73 \sqrt{1-4 \nu }+73 \big) \big(\hat{L}.\hat{S}^{}_2 \big) \chi _2\big]\,.
\end{split}
\end{equation}

\begin{subequations}
\label{Eq.Hparams}
\begin{align}
A^{(i)}_{5} \equiv &{} \left(\frac{M^{}_i}{M}\right)^3 \left(\hat{L}.\hat{S}^{}_i\right)\chi _i \left(3 \chi_i^2+1\right)\,,\\
A^{(i)}_{8} \equiv &{} ~4 \pi A^{(i)}_{5}+\left(\frac{M_i}{M}\right)^4 \left(3 \chi_i^2+1\right)\nonumber \\
&\quad\quad\quad\quad\quad\quad\quad \times \left(\sqrt{1-\chi_i^2}+1\right)\,.
\end{align}
\end{subequations}

These expressions will be used later to calculate the dephasing which will lead us to understand the observability of the area quantization.

\section{Observability before plunge}

In the previous section, we have discussed the impact of area quantization on the inspiral phase and how they can be modeled. In this section, we will focus more on the details of the observability of this phenomena.

\begin{figure}
\centering
\includegraphics[width=8.0cm]{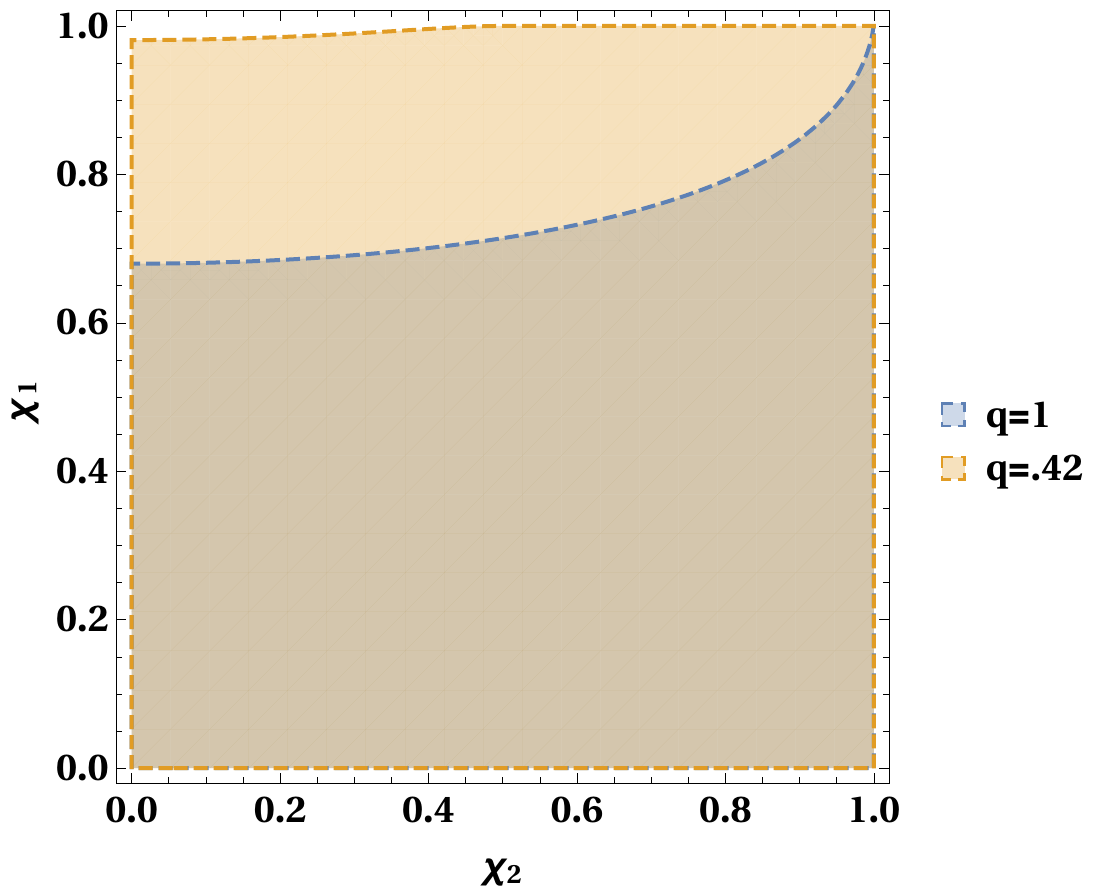}
\caption{We plot $f_0 < f_{contact}$ region with respect to the individual spins of the binary components for mass ratio $q=1$ in blue and $q=.42$ in orange. The dashed lines show the boundary. Note, we are focusing only on the observability of the first body's quantum nature. This is why the plot is asymmetric between $\chi_1$ and $\chi_2$ even for $q=1$. A similar kind of result can be found if we focus on the observability of the second body's quantum nature.}
\label{fig:f0_comparison}
\end{figure}

\subsection{Absorption frequency vs contact frequency}

For a given spin $\chi$ and area separation $\alpha$, there is a lowest absorption frequency, i.e. $\omega_0 =2\Omega_H$. Only above this frequency, TH will be present for QBH.
The range of frequency covered in an inspiral upto merging of the two bodies in a binary is bounded from above by the frequency corresponding to the length scale of the radial surface position of the bodies. In a binary, the closest reach in the inspiral is $r_{H1}+r_{H2}$, where they touch each other. Using Kepler's third law, corresponding frequency to this length can be found as, 

\begin{equation}
f_{contact} = \frac{1}{\pi M_1} \sqrt{\frac{1+q}{(1+\sqrt{1-\chi_1^2} + q + q\sqrt{1-\chi_2^2})^3}},
\end{equation}
where $q=M_2/M_1$.

Note that both of the frequencies $(f_0, f_{contact})$ are functions of mass and spin. Where $f_0$ is the lowest frequency of the TH lines for QBH and $f_{contact}$ is the upper frequency before they merge, above which TH phenomenon is not relevant. Hence, non-zero discretized structure of TH flux can only be observed iff $f_0 \leq f_{contact}$. In Fig. \ref{fig:f0_comparison}, we plot $f_0<f_{contact}$ region with respect to the individual spins of the binary components for mass ratio $q=1$ in blue and $q=.42$ in orange. The dashed lines show the boundary. Note, we are focusing only on the observability of the first body's quantum nature (i.e. $f_0$ of the first body). This is why the plot is asymmetric between $\chi_1$ and $\chi_2$, even for $q=1$. A similar kind of result can be found if we focus on the observability of the second body's quantum nature.

From the Fig. \ref{fig:f0_comparison}, we notice that for a given value of $q$ there is a region above the dashed curve where $f_0 / f_{contact}<1$ is not satisfied. We find that below $\chi_1=0.679667$, the condition is always satisfied. However, above this spin value the allowed region satisfying $f_0 <f_{contact}$ depends on both $q$ and $\chi_2$. But for $q<.384646$, for entire range of $\chi_1$ and $\chi_2$ the condition $f_0 < f_{contact}$ will be satisfied. In these ranges where $f_0<f_{contact}$, non-zero TH will be present during the inspiral. But the feature of it will be completely different from TH of CBH, as discussed before. This, as a result, can be used as a signature of the area quantization. In the ranges where $f_0 > f_{contact}$ TH will be exactly equal to zero for QBH. In the parameter range where TH is expected to be absent for QBH, a model selection strategy with the Bayes factor can be used to test for the presence of ``classical TH" to distinguish between CBH and QBH. It has been addressed in Ref. \cite{Datta:2020gem}. Hence, using TH as a possible probe for area quantization seems viable.

\begin{figure}
\centering
\includegraphics[width=\linewidth]{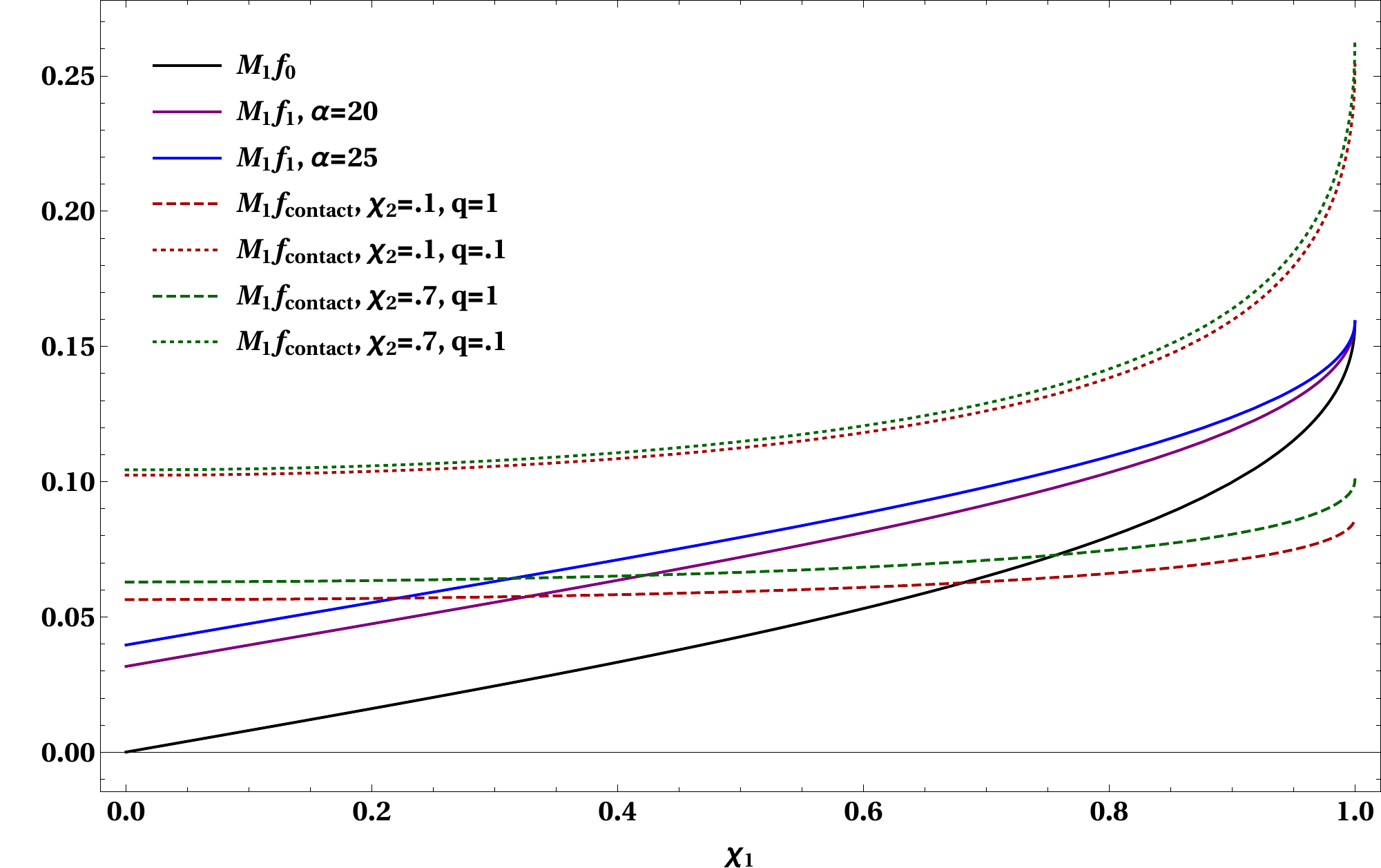}
\caption{We plot first body's $f_0$ and $f_1$ for $\alpha =20, 25$ along with the contact frequencies for different parameter values of the binary.}
\label{fig:spin_bound}
\end{figure}

To measure $\alpha$, it is necessary to measure at least one higher frequency absorption line, namely $f_1$. This implies that $\alpha$ will be measurable if $f_1 < f_{contact}$. In Fig. \ref{fig:spin_bound}, we plot first body's $f_0$ and $f_1$ for $\alpha =20, 25$ along with the contact frequencies for different parameter values of the binary. For equal mass binaries $(q=1)$ (dashed red and green line), $f_0$ is greater than $f_{contact}$ for higher spin of the first body. This is in accordance with Fig. \ref{fig:f0_comparison} and illustrates the argument described in earlier paragraphs. In Fig. \ref{fig:spin_bound}, we observe that for smaller $q$, $f_0 < f_{contact}$ for all spin values of the first body (dotted green and red line).

\begin{figure}
\centering
\includegraphics[width=\linewidth]{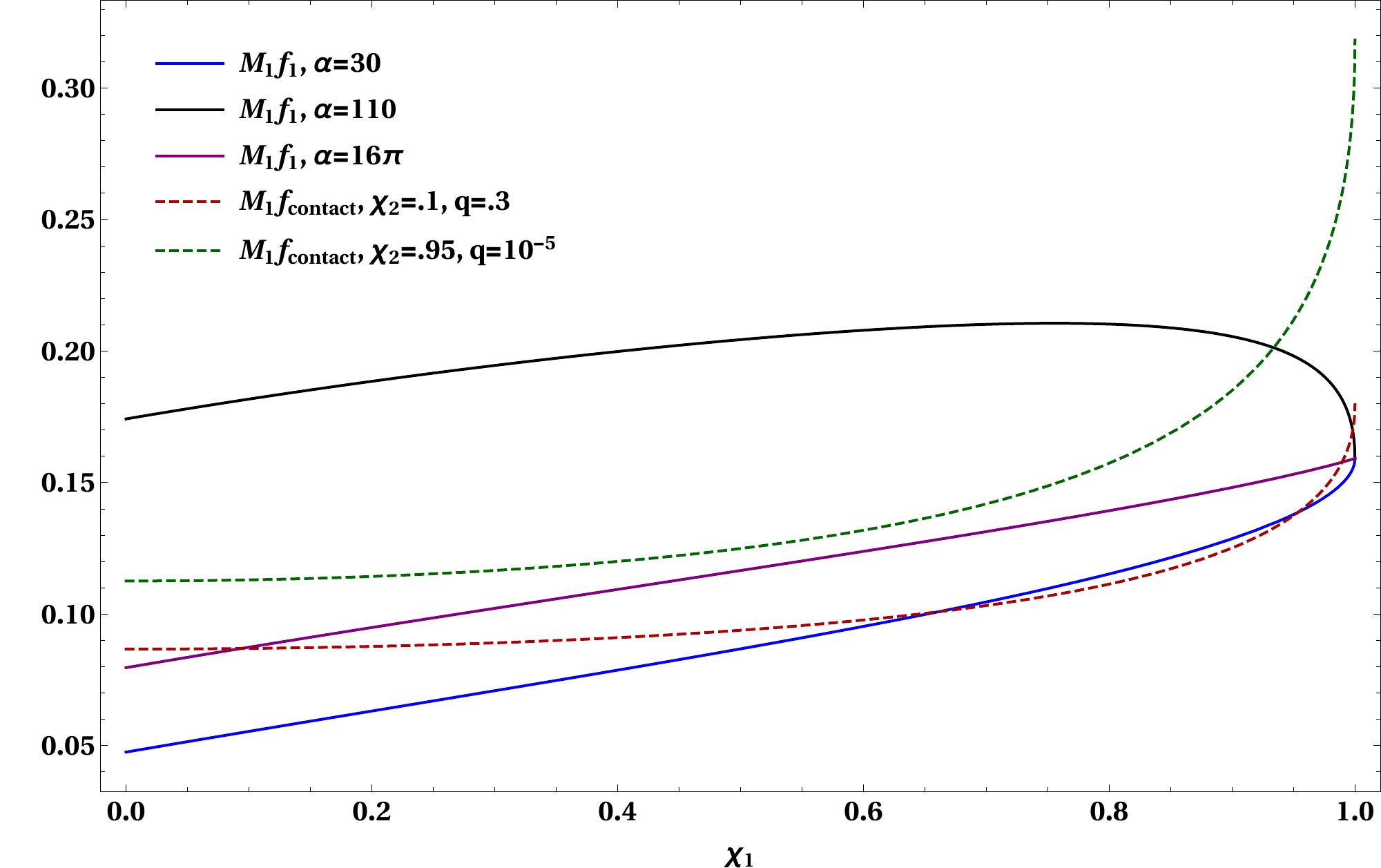}
\caption{We plot different contact frequencies of the binaries for $\chi_2=.1,\,q=.3$ in red dashed curve, and $\chi_2=.95,\,q=10^{-5}$ in green dashed curve. We also plot $f_1$ for $\alpha = 30,\,110,\,16\pi$ in blue black, and purple.}
\label{fig:spin_bound2}
\end{figure}

Similar behavior is shown by the $f_1$ curves (solid blue and purple curves). But in this case the value of $\alpha$ becomes an important parameter. By increasing $\alpha$, $f_1$ curves can be shifted upwards, as has been illustrated by the blue curve ($\alpha = 25$)
having higher values than the purple curve ($\alpha = 20$). Therefore, in the near-equal mass binaries beyond a maximum value of $\alpha$, say $\alpha_{max}$, $f_1 > f_{contact}$ for all $\chi_1$. Hence, even if area quantization is present in nature, we can only measure it using TH if and only if $\alpha<\alpha_{max} (\chi_1, \chi_2, q)$. This can be understood by noting that the red dashed curve intersects with the $\alpha=25$ (solid blue) curve at a smaller $\chi_1$ value than the $\alpha=20$ (solid purple) curve. It also implies that $\alpha_{max}$ will be a function of the binary parameters.

\begin{figure*}
\includegraphics[width=8.8cm]{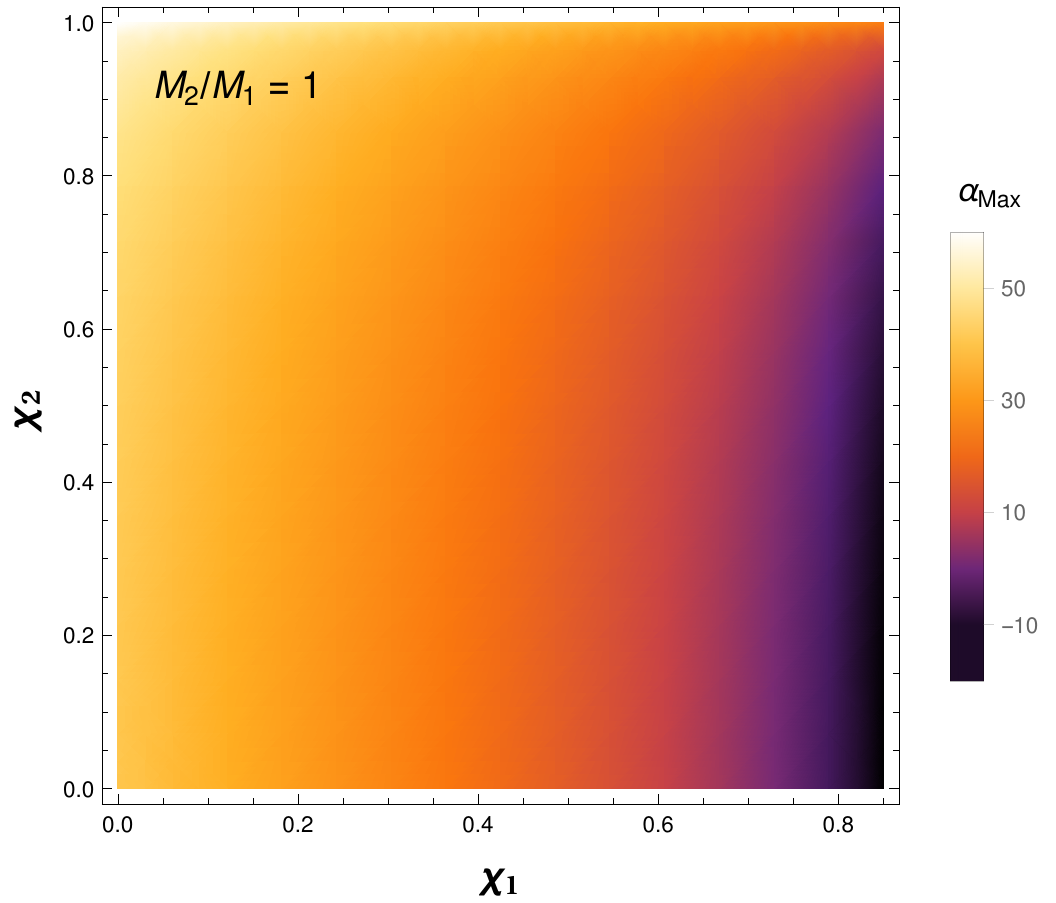}
\includegraphics[width=8.8cm]{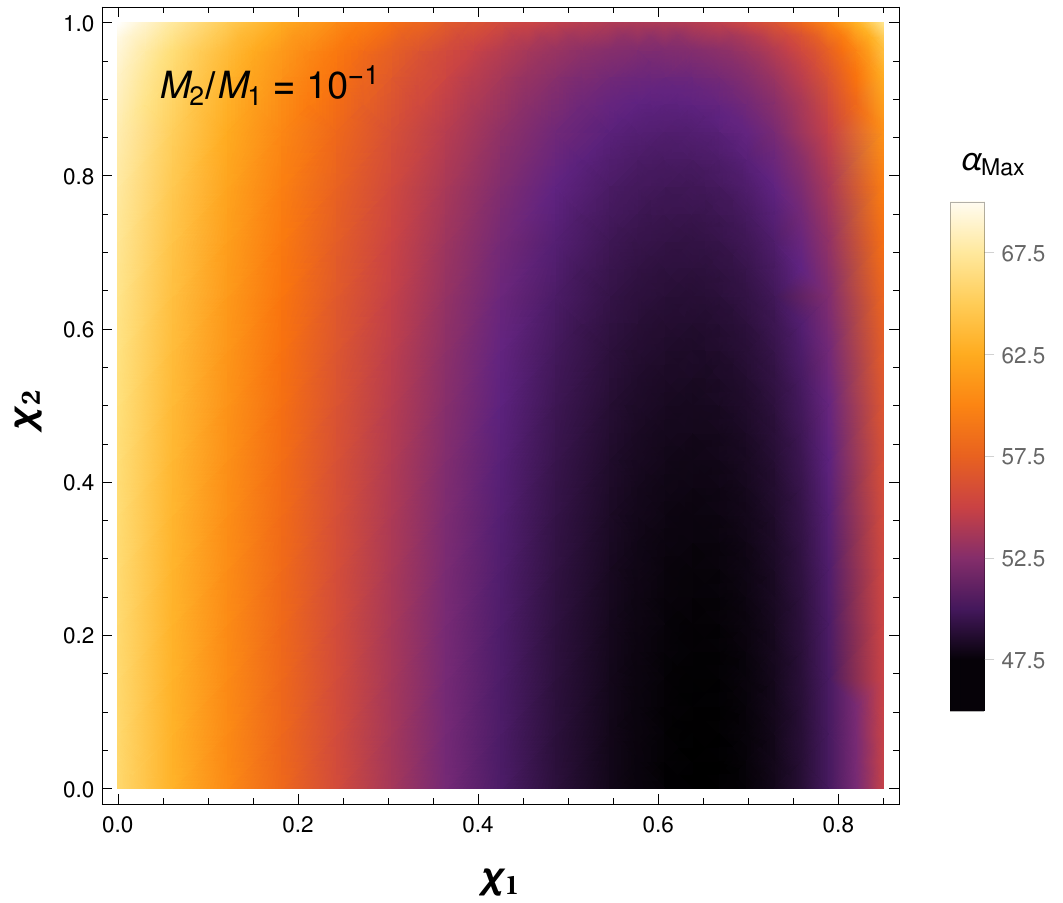}\\
\includegraphics[width=8.8cm]{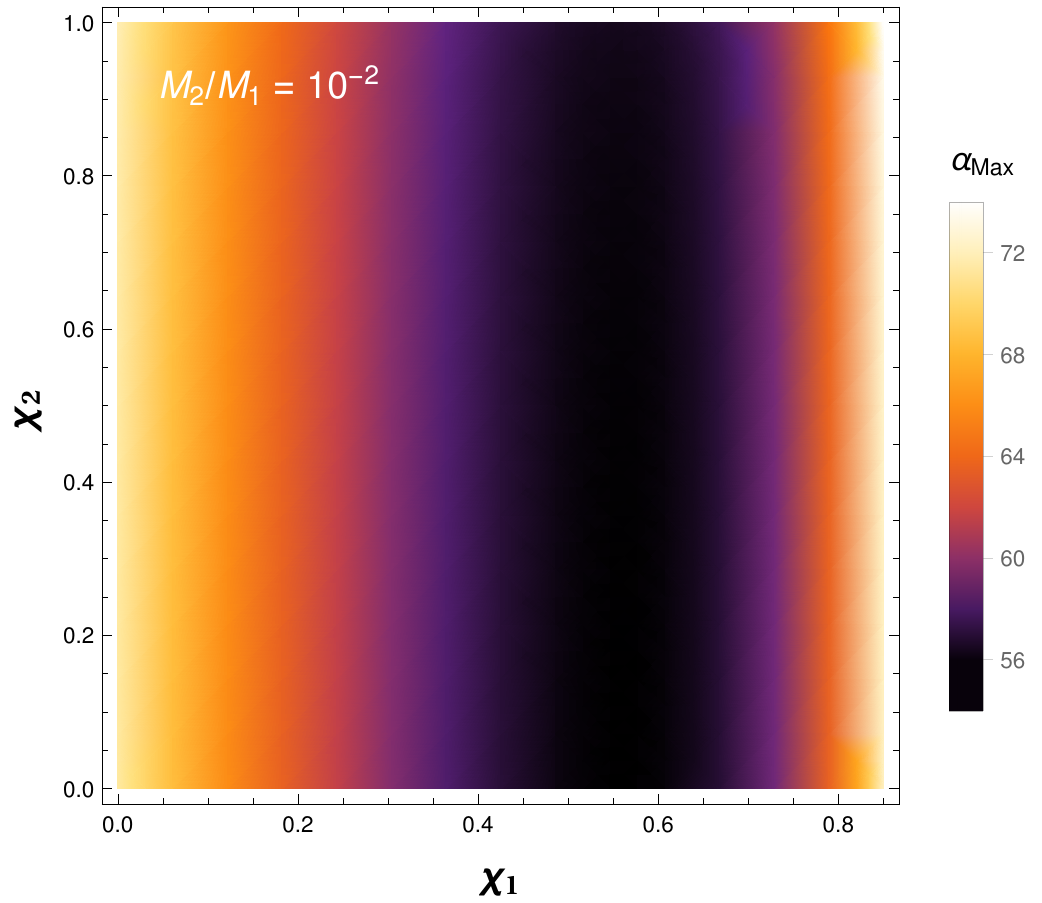}
\includegraphics[width=8.8cm]{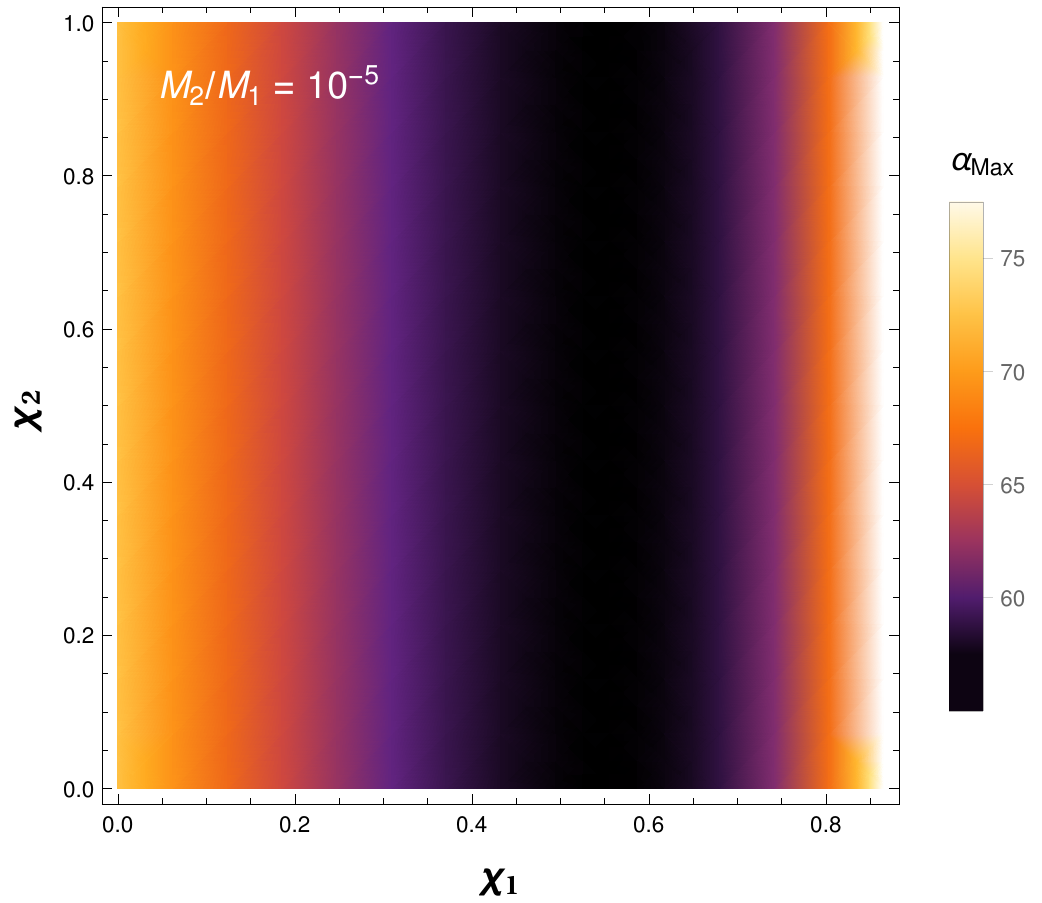}
\caption{We plot the $\alpha_{max}$ above which $f_1$ will not be observable before the binary merges. Since the comparison arises due to the $f_{contact}$, it depends on $\chi_2$ and the mass ratio $q=\frac{M_2}{M_1}$. $\alpha_{max}$ has been plotted in the colour bar w.r.t. $\chi_1$ and $\chi_2$.}\label{fig:alpha_max}
\end{figure*}

We calculate $\alpha_{max}$ by using $f_1 = f_{contact}$. We find,
\begin{equation}
\begin{split}
\frac{\alpha_{max}}{32\pi} =& \frac{1+\sqrt{1-\chi_1^2}}{\sqrt{1-\chi_1^2}}\left[\frac{1+q}{(1+q+\sqrt{1-\chi_1^2} + q\sqrt{1-\chi_2^2})^3}\right]^{\frac{1}{2}}\\
-& \frac{\chi_1}{ 2\sqrt{1-\chi_1^2}}.
\end{split}
\end{equation}
In Fig. \ref{fig:alpha_max}, we show the value of $\alpha_{max}$ for different binary parameters. As can be seen in the plot in the upper left corner, for high values of $\chi_1$ in equal mass binaries $(q=1)$, $\alpha_{max}<0$. Which in accordance with the Fig. \ref{fig:spin_bound}. It implies that in equal mass binaries there will be no observable higher absorption lines if the components of the binaries are spinning very fast. In equal mass binaries we find the maximum observable $\alpha_{max}\sim 54$. If $\alpha$ is very high $(\mathcal{O}(10))$ in nature, then to measure it the spin of the corresponding body in the equal mass binary requires to be small. In the figures for $M_1>M_2$ we restrict ourselves with $\chi_1 \leq .85$, primarily because beyond this value $\alpha_{max}$ can take extremely large values, i.e. $\alpha_{max} > 600$ for $\chi_1\sim .998$.

\begin{figure}
\centering
\includegraphics[width=.8\linewidth]{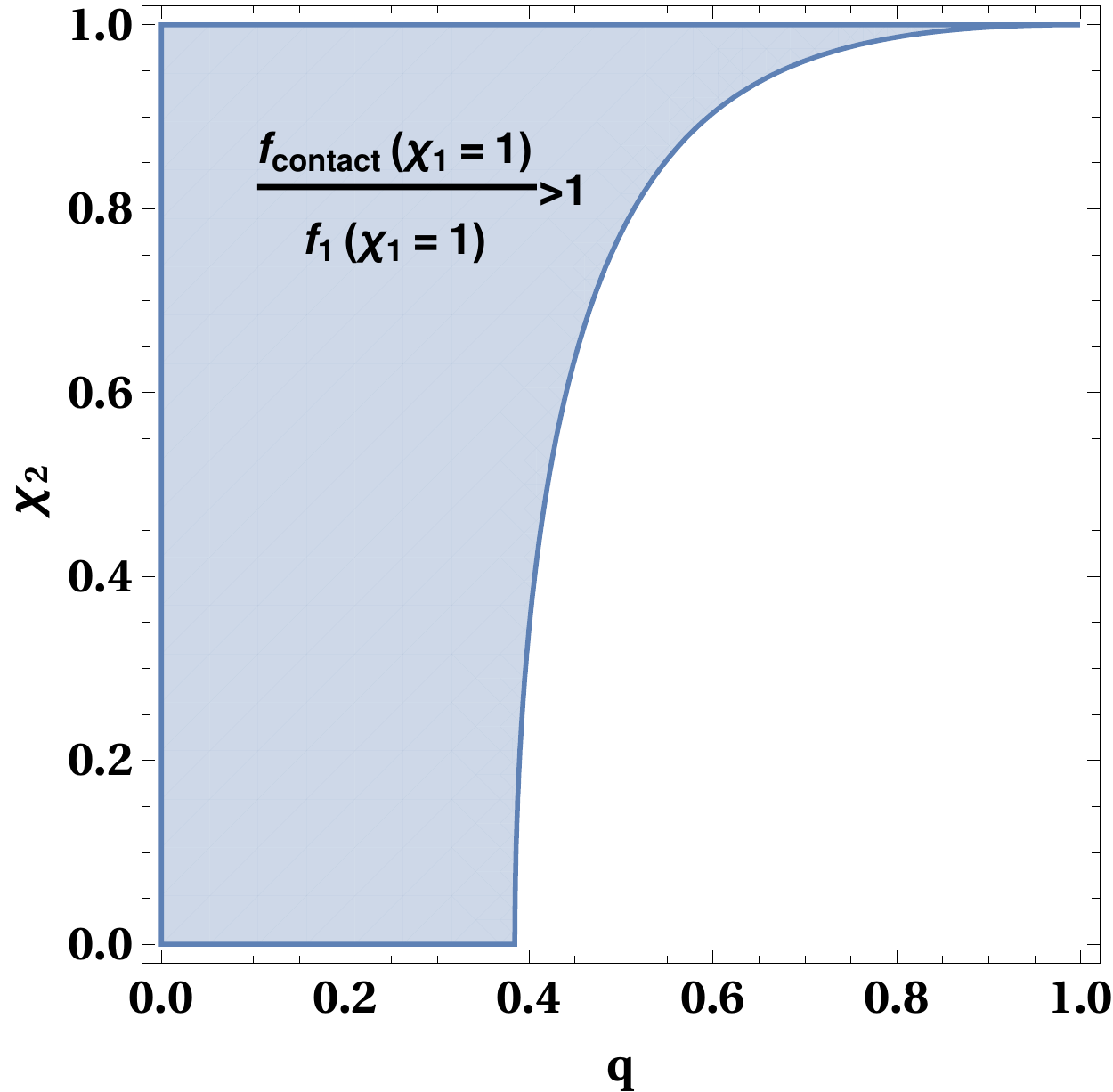}
\caption{The blue shaded region represents the parameter region that satisfies $f_1(\chi_1=1)<f_{contact}(\chi_1=1)$. We find for $q\leq 0.384646$ this condition is satisfied for the entire range of $\chi_2$.}
\label{fig:alphaMax_region}
\end{figure}

Note, for $M_1 > M_2$, $\alpha_{max}$ reaches the minimal values in the mid range of $\chi_1$ and grows after that. This can be understood from Fig. \ref{fig:spin_bound2}. Dashed red and green curves are the $M_1f_{contact}$ for {$q=.3,\,\chi_2 =.1$} and {$q=10^{-5},\,\chi_2 =.95$}, respectively. Blue, black, and purple curves represents $f_1$ for $\alpha = 30,\, 110,\, 16\pi$ respectively. We find that the $M_1f_{contact}$ frequency curves can intersect $M_1f_1$ curves at two points $\chi_1 = \chi_{low},\, \chi_{high}$. This creates an intermediate region of $\chi$ where the $\alpha_{max}$ decreases then it increases again for high spin values.

The reason for such behavior is connected with the different behavior of $M_1f_{contact}$ and $M_1f_1$ as a function of $\chi_1$. $f_{contact}$ is a monotonically increasing function with respect to spin. But $M_1f_1$ can have a maxima for a range of $\alpha$. This is primarily because $M_1f_1$ is fixed to the value of $1/2\pi$ when $\chi_1=1$ , which is independent of $\alpha$. But $M_1f_{contact}$ at $\chi_1=1$, increases with increasing $\chi_2$ and decreasing $q$.

If $f_1(\chi_1=1)=1/(2\pi M_1)<f_{contact}(\chi_1=1)$ then even for arbitrarily large values of $\alpha$ there will be at least one intersection point between $f_1$ and $f_{contact}$ near $\chi_1=1$. In Fig. \ref{fig:spin_bound2}, this phenomena is being exhibited by $\alpha=110$ curve. In Fig. \ref{fig:alphaMax_region}, we have plotted the region that satisfies $f_1(\chi_1=1)<f_{contact}(\chi_1=1)$. We find for $q\leq 0.384646$ this condition is satisfied for the entire range of $\chi_2$.

Note, in Fig. \ref{fig:spin_bound2}, $\alpha=110$ curve has a maxima while there are no maximas present for lower values of $\alpha$ shown in Fig. \ref{fig:spin_bound}. For a fixed $\alpha$ the condition for extrema of $f_1$ is $f_1'(\chi_1) =0$, where prime denotes the derivative with respect to $\chi_1$. This results as,

\begin{equation}\label{eq:alphaMax}
\alpha = \frac{16\pi (1+\sqrt{1-\chi_1^2})}{\chi_1}.
\end{equation}
It is easy to verify that the right hand side of Eq. (\ref{eq:alphaMax}) has a minimum value of $16\pi$. Hence for $\alpha<16\pi$ this condition will not be satisfied and there will not be any extrema. Interestingly, in the literature the range of expected values is $1<\alpha<30$ \cite{Cardoso:2019apo}. Therefore in this physical range maxima will never occur. Hence, Fig. \ref{fig:alpha_max} implies that for $q\leq.1$ entire physical range can be probed. This can help us in putting bound on $\alpha$ from observation.

\begin{figure}[ht!]
\centering
\includegraphics[width=\linewidth]{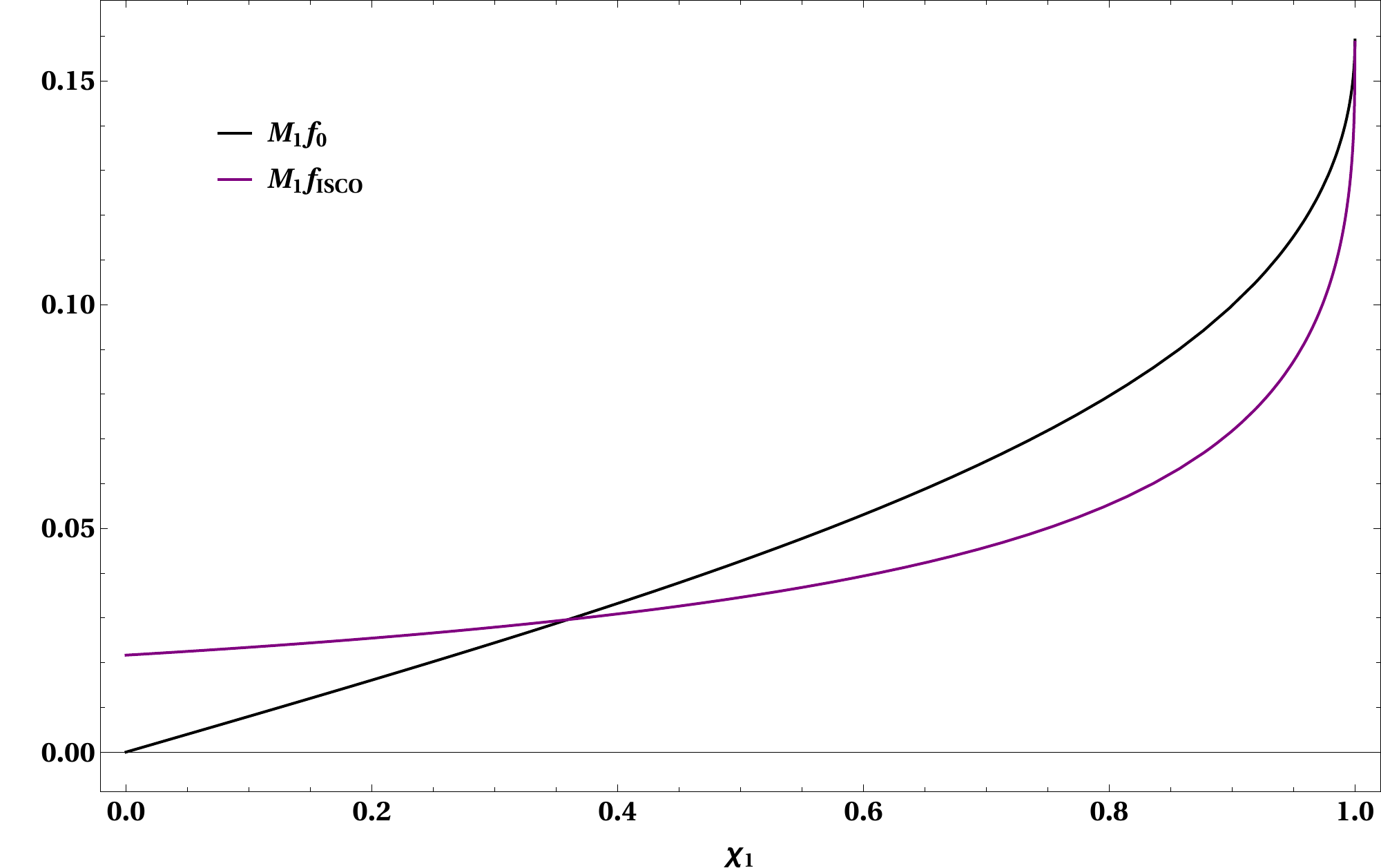}
\caption{We plot $f_0$ and ISCO frequency in EMRI. Above $\chi = .359405$,  $f_{\rm ISCO} < f_0$. As a result, above this spin $\delta \phi = 0$ and below this spin value $\delta \phi \sim \mathcal{O}(.1)$. For CBH $\delta\phi \sim \mathcal{O}(10^2)$ and for $\chi = .9$, $\delta \phi \sim \mathcal{O}(10^3)$ \cite{ Datta:2019epe}. \label{fig:isco}}
\label{fig:phase}
\end{figure}

\subsection{Dephasing in the inspiral phase}

Using Eq. (\ref{eq:phase correction}), we calculate the dephasing in radian due to TH in QBH. A useful estimator to describe the effects of  $\alpha$ in the phase is the total number of GW cycles $(\equiv N)$ that accumulates within a given frequency band of the detectors. In terms of the frequency-domain phase $\Psi(f)$ this can be defined as,

\begin{equation}
\label{eq:N}
N \equiv \frac{1}{2\pi}\int_{f_{min}}^{f_{max}} f df \left(\frac{d^2\Psi(f)}{df^2}\right).
\end{equation}
By substituting Eq. (\ref{eq:phase correction}) into Eq. (\ref{eq:N}) the relative number of GW cycles $(\equiv \Delta N)$ that is contributed by this effect and accumulated within the frequency band $f \in [ f_{min} , f_{max} ]$  is calculated. We find only in case of extreme mass ratio inspiral (EMRI) dephasing is reasonable. For EMRI we take $f_{min}=.5\, {\rm mHz},\,\,f_{max} = f_{ISCO}$. We compute the dephasing (in radian), $\delta \phi = 2\pi\Delta N$, by taking into account of non-zero $\alpha$. We find that above $\chi = .359405$, $\delta\phi = 0$ and below it $\delta\phi \sim \mathcal{O}(10^{-1})$ even for high values of $\alpha$ (in the physical range).

\begin{figure}
\includegraphics[width=8.8cm]{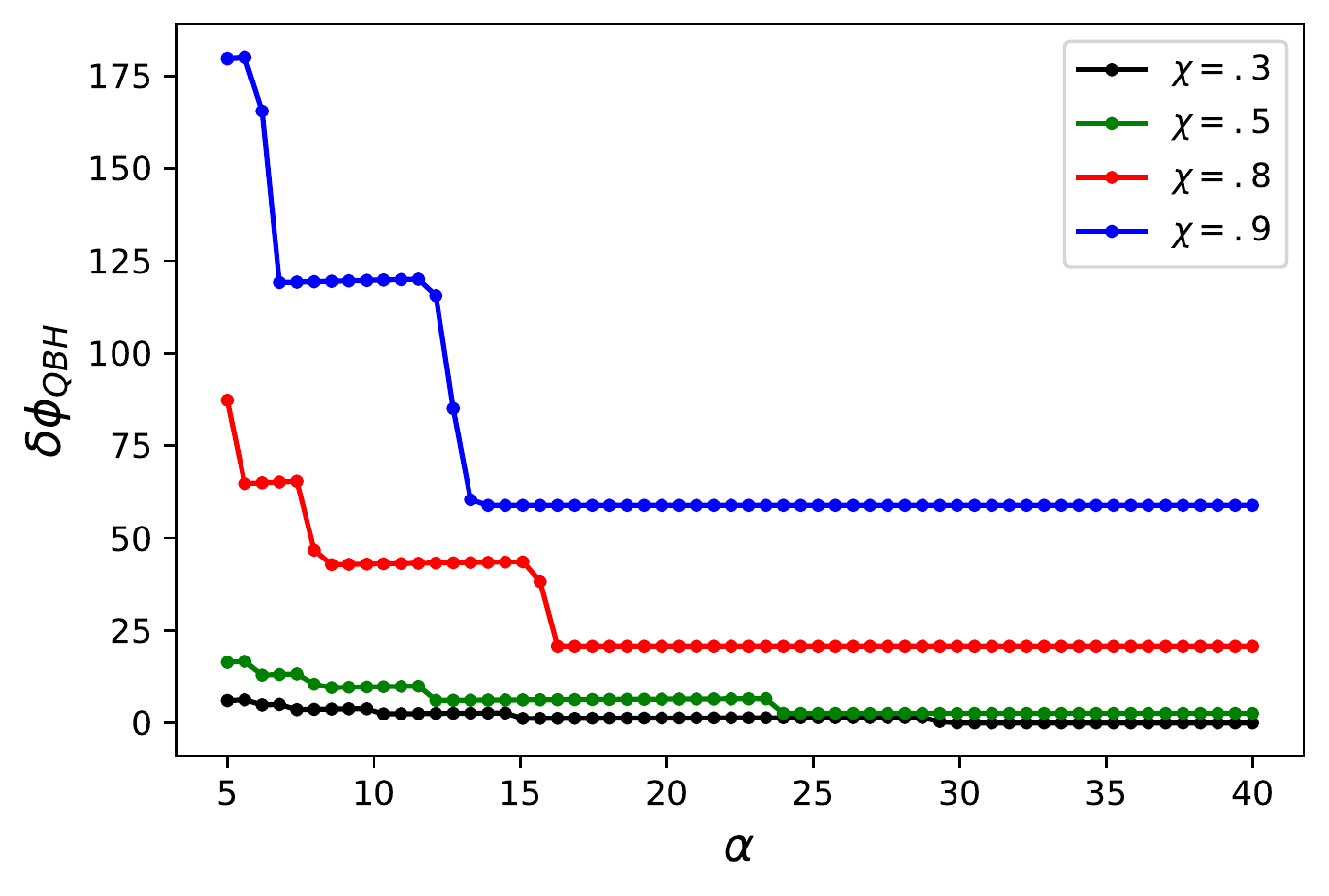}
\includegraphics[width=8.8cm]{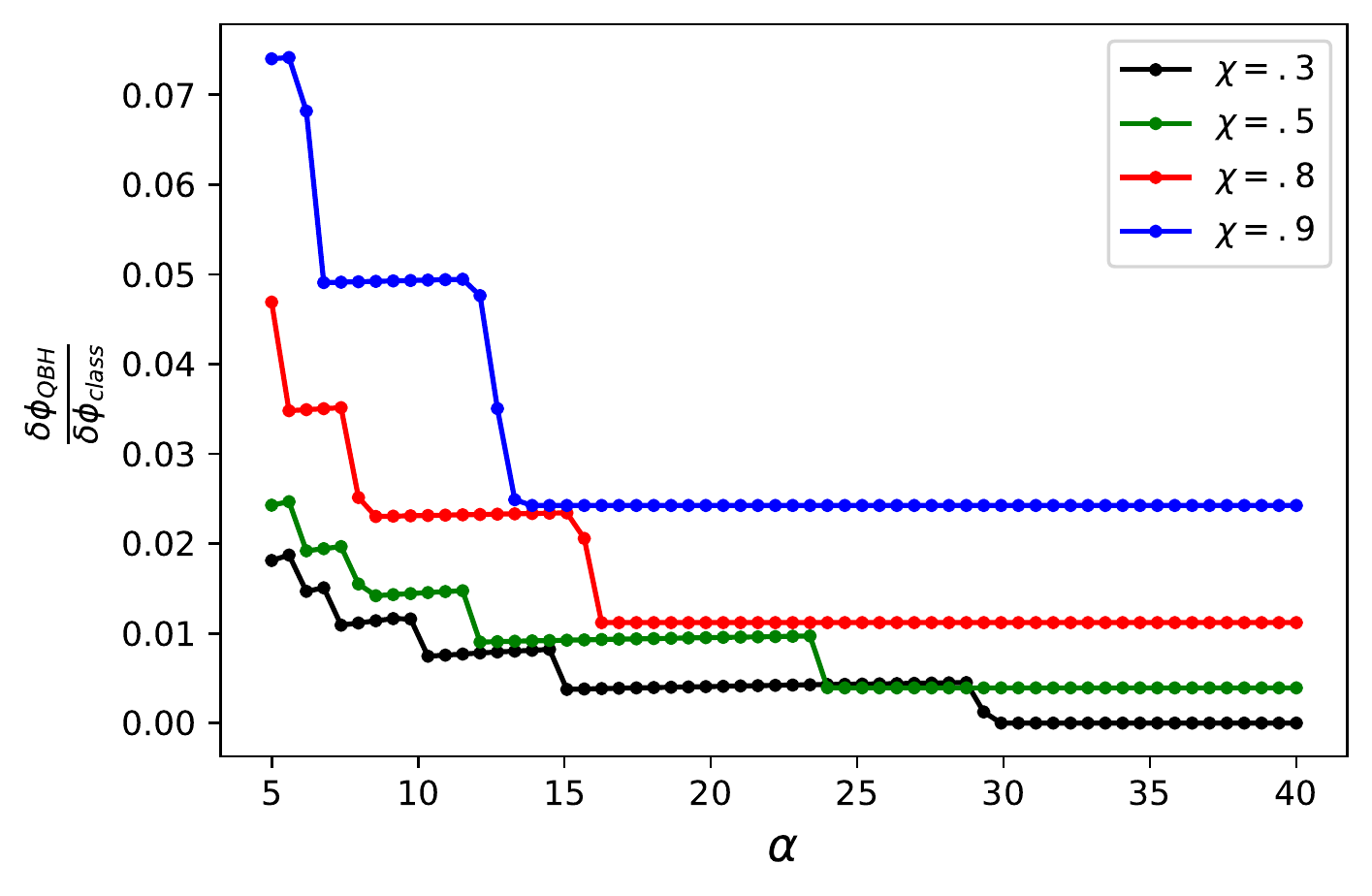}
\caption{The dephasing (in radian) from ISCO frequency to the light ring frequency is calculated. In the upper panel, the dephasing for QBH is demonstrated with $\alpha$. In the lower panel, the ratio of QBH and CBH dephasing is demonstrated in the same frequency band.}\label{fig:dephasing}
\end{figure}

This can be understood from Fig. \ref{fig:isco}. In Fig. \ref{fig:isco} we plot $f_0$ and the frequency corresponding to the inner most stable circular orbit (ISCO) of EMRI, with respect to $\chi$. In the figure above $\chi = .359405$, $f_0>f_{ISCO}$. In such cases there will be no tidal heating contribution for QBH. This is a telltale signature, as for CBH in EMRI  $\delta \phi_{CBH} \sim \mathcal{O}(10^2)$($\mathcal{O}(10^3)$)for $\chi = .3$(.9). Therefore, it can safely be said that absence of any TH is a potential signature of area quantization. But unfortunately, if the bodys are spinning very fast then except for the absence of TH it is not possible to find any other feature of area quantization in the inspiral up to ISCO. As a result, it can potentially lead to degeneracy between horizonless ECOs and QBHs. But this degeneracy can get broken in the merger phase. This has been discussed in the next section.

\subsection{Dephasing in the merger phase}

In the previous section, we demonstrated that the TH effect of QBH is vanishingly small in the inspiral phase of a binary. As a result, the dephasing due to TH in QBH is either vanishes or is vanishingly small. However, our investigation suggests that $f_0$ is always smaller than $f_{contact}$ in EMRI, implying it is possible to find signatures in the merger and post-postmerger phase. To explore it, we again calculate dephasing $(\delta\phi)$ for an EMRI with a $10^6M_{\odot}$ supermassive body and a $10M_{\odot}$ companion. But this time we integrate Eq. (\ref{eq:N}) from ISCO frequency to the frequency corresponding to the light ring (see Ref. \cite{Taracchini:2013wfa} for definition).

In the upper panle of Fig. \ref{fig:dephasing} we plot the accumulated dephasing due to QBH $(\delta\phi_{QBH})$ in the frequency band, w.r.t. $\alpha$. In lower panel we plot the ratio of dephasing between QBH TH and CBH TH, namely $(\delta\phi_{QBH}/\delta\phi_{class})$. As expected, the dephasing is higher for higher spin values. We also find that with increasing $\alpha$, the dephasing decreases but also stays roughly unchanged in a range of $\alpha$, creating a stair-like structure. The stair structure originates due to the different number of absorption lines present in the frequency band. With increasing $\alpha$, the lines get more separated. Due to the increased separation, if a line goes out of the frequency band then the dephasing drops. Otherwise, it increases very slowly. The increase occurs due to the increase in the strength of TH with the absorption lines shifting towards higher frequency while staying in the frequency band. It implies that the corresponding dephasing in EMRI can be used to measure $\alpha$ and $\Gamma$ in the merger phase. However, we have used a stationary phase approximation to find this dephasing, which is unlikely to be applicable. Hence, it remains to be seen what happens when realistic phasing is considered in the merger phase. However, the figures establish that the merger phase can introduce significant deviation from CBH contribution, resulting in the measurability of $\alpha$ and $\Gamma$.

\section{Probing Hawking radiation}

We have explicitly shown that the presence of area quantization affect the tidal heating phenomena of a QBH in a binary. Since, tidal heating affects the inspiral rate of a binary, imprints of area quantization is possible to find in the emitted GW from the system. This phenomena manifests through two parameters, quantization spacing $(\alpha)$ and the line width $(\Gamma)$. We have explored in details the effect of $\alpha$ and $\Gamma$ in the GW.

The width $\Gamma$ of the energy levels is written as the inverse of a decay rate, $\tau$ , as $\Gamma = \hbar/\tau $. This timescale is associated with the spontaneous decay of the BH energy states due to Hawking radiation, and has been estimated as $\tau \equiv -\frac{\langle\hbar\omega\rangle}{\Dot{M}}$, where
$\langle\hbar\omega\rangle$ denotes the average frequency over all the decay channels, and $\Dot{M}$ is the power (check Ref. \cite{Agullo:2020hxe} for more details).

During observation it is possible to treat $\Gamma$ as an independent parameter, rather than using Eq. (\ref{eq:gamma}). Since we will measure spin $\chi$ along with $\Gamma$, the measured value of $\chi$ can be used to find a fit between $\Gamma$ and $\chi$ using data from multiple observations. Once such fitting function is available, it is possible to compare the observed fitting function with the analytic expression in Eq. (\ref{eq:gamma}). Any deviation from Eq. (\ref{eq:gamma}) will represent the limitation of the estimation of $\tau$. This deviation can arise due to modification of Hawking radiation itself or due to some beyond standard model channels not considered in Ref. \cite{Agullo:2020hxe}.

As a result, we will have observational information regarding Hawking radiation as well as the nature of the quantum states of QBH. This therefore, has the potential to shed light on the information loss paradox from the side of the observation \footnote{From inspiraling binary, upper bound on the Hawking temperature has been found in Ref. \cite{Chung:2020uqj}. However this value is very high with respect to the expected value of the Hawking temperature of these systems.}. Note, it is unlikely that TH can be used for that purpose unless the spin of the bodies are  low and the signal to noise ratio (SNR) is very high. But probing Hawking radiation by measuring $\Gamma$, as described in previous paragraph, will be valid even in merger and postmerger phase.

\section{Observability}

In our work, we establish that due to the presence of area quantization, effect of TH becomes vanishingly small in EMRI. Above $\chi \sim .36$ it results in vanishing TH. Therefore, measuring no TH in EMRI will be an indication towards area quantization, this aspect was proposed in Ref. \cite{Agullo:2020hxe}. We agree with their claim after a detailed investigation. However, as TH is not exactly equal to zero below $\chi \sim .36$, with high SNR, $\alpha$ and $\Gamma$ can be measured. This will be investigated in recent future. Another key point is, we have not included the effect of resonances which can modify our conclusion.

Although for high spins, TH vanishes exactly below ISCO frequency, $f_0$ and $f_1$ are less than contact frequency. This implies that the frequency corresponding to the area quantization will be in the range of the merger and postmerger phase frequencies. As a result, in the merger phase of an EMRI it will lead to non-zero dephasing for any non-zero spin of the supermassive body.

The vanishing of TH for high spins below ISCO implies that there will not be any absorption of GW flux in the inspiral phase. As a result, this can lead to ergo region instability \cite{Maggio:2017ivp, Maggio:2018ivz}, which can have observable consequences. This requires a detailed investigation. Note, that the information of area quantization is solely captured by the first term in Eq. (\ref{eq:base equation omega}), while the second term corresponds to angular momentum states of the QBH. For $\ell=m=2$ mode the second term is similar to the $m\Omega_H$ that connects the frequency at the far range to the frequency at the near horizon range. This indicates that even for a different area quantization law only the first term in Eq. (\ref{eq:base equation omega}) will get modified while the second term will stay intact for $\ell=m=2$ mode (assuming Eq. (\ref{eq:classical mass}) stays valid). In such a case for nonzero $\chi$ there will always be a minimum frequency $\propto \Omega_H$ below which TH vanishes. Therefore non-vanishing TH can be a strong indication of a violation of any kind of area quantization.

\section{Discussion}

In this work, we have explored the possibility of observing BH area quantization using tidal heating. We have argued that it can be done by assuming a frequency-dependent reflectivity of the QBH. We have explicitly constructed a model for such reflectivity. Once implemented, we find as expected, the tidal heating flux becomes quantized, unlike classical black holes. Using the quantized structure of the tidal heating flux, we find the expression of the phase of the QBH binary. Since in the presence of area quantization, TH is absent below $f_0$, it will be hard to distinguish between QBH and horizonless ECOs for $f<f_0$. To address it, we explored the parameter range where $f_0<f_{contact}$ is satisfied. We find that below mass ratio $q<.384646$, $f_0<f_{contact}$ is always satisfied. We have also found the expression for $\alpha_{max}$ for a set of binary parameters. We find that for equal mass binary observable $\alpha_{max}<54$. For binaries with $q<.1$ the entire physical range of $\alpha$ is observable. By calculating dephasing, we demonstrate that in EMRI the effect of area quantization is very strong in the sense that the effect of TH vanishes for higher spin values below ISCO. This is drastically different from CBH, where TH can add several cycles in the phase. However, in the merger phase of an EMRI area quantization leads to non-zero dephasing. We also discuss the possibility of probing Hawking radiation with the measurement of $\Gamma$.

To model the lines we have used the Hann window function. They can also be modeled with Lorentzian profile, 
\begin{equation}
\mathfrak{P}(\omega) \propto  \frac{2}{\pi}\frac{\Gamma}{4(\omega - \omega_n)^2 + \Gamma^2}.
\end{equation}
Under such modeling also the conclusion stays the same (check Ref. \cite{Chakravarti:2021jbv} too). However, there remain differences in the interpretation of the different modeling. I.e., if the normalized Lorentzian or Gaussian profiles are used then at the position of the absorption frequency $f_i$, the maximum value of each profile can be larger than one. This, as a result, is different from the Hann window function. In choosing the Hann function, we kept the maximums of $\mathcal{H}$ at equal to 1. It implies that when absorption is allowed, it will be like CBH at these frequencies. In Gaussian or Lorentzian profile $\mathcal{H}(f_i)>1$, which is different from CBH. It requires an investigation from the theoretical side.

In Ref. \cite{Agullo:2020hxe} it has been assumed that the QBH can emit at any real frequency $\omega$ during the Hawking evaporation process. It will not be true if the energy levels are quantized. It is because only such decay channels will be allowed that end at an allowed energy level. If that is taken into account while calculating the values of $\Gamma$, not only the values of $\Gamma$ will decrease, but also it will depend on the quantization $\alpha$.

It is extraordinary that we are at a juncture when Planck scale physics can be tested with the current and the upcoming detectors. We have shown that the tidal heating in the inspiral phase of a binary will look quite different from binary CBHs. The area quantization of QBHs can lead to peculiar features in tidal heating flux, which for CBH is a 2.5 PN$\times\log v$ effect in the leading order.
The analysis presented here heavily depends on Bekenstein-Mukhanov's semi-heuristic arguments on quantum BHs. It is mind-boggling that the quantum aspects of BHs are within the reach of observations. It is high time to advance the quantum gravity frameworks so that concrete predictions are possible to make.

\section*{Acknowledgement}
It is a pleasure to thank Adri$\acute{\rm a}$n del Rio and Vitor Cardoso for providing us numerical data that has been used to find the analytical expression of the line width. We thank Adri$\acute{\rm a}$n del Rio and Vitor Cardoso for reading an earlier draft of the paper carefully and giving us valuable suggestions.
We thank Yash Bhargava, Bhaskar Biswas, and Soumak Maitra for useful discussions.
SD would like to thank University Grants Commission (UGC), India, for financial support for a senior research fellowship.
KSP acknowledges the support of the Science and Engineering Research Board (SERB), India, and the Netherlands  Organisation for  Scientific  Research (NWO). This work was done with partial support provided by the Tata Trusts. This paper has been assigned LIGO Document Number LIGO-P2000115. 

{\it We would like to thank all of the essential workers who put their health at their risk during the COVID-19 pandemic, without whom we would not have been able to complete this work.}

\bibliography{bib.bib}

\begin{thebibliography}{65}%
\makeatletter
\providecommand \@ifxundefined [1]{%
 \@ifx{#1\undefined}
}%
\providecommand \@ifnum [1]{%
 \ifnum #1\expandafter \@firstoftwo
 \else \expandafter \@secondoftwo
 \fi
}%
\providecommand \@ifx [1]{%
 \ifx #1\expandafter \@firstoftwo
 \else \expandafter \@secondoftwo
 \fi
}%
\providecommand \natexlab [1]{#1}%
\providecommand \enquote  [1]{``#1''}%
\providecommand \bibnamefont  [1]{#1}%
\providecommand \bibfnamefont [1]{#1}%
\providecommand \citenamefont [1]{#1}%
\providecommand \href@noop [0]{\@secondoftwo}%
\providecommand \href [0]{\begingroup \@sanitize@url \@href}%
\providecommand \@href[1]{\@@startlink{#1}\@@href}%
\providecommand \@@href[1]{\endgroup#1\@@endlink}%
\providecommand \@sanitize@url [0]{\catcode `\\12\catcode `\$12\catcode
  `\&12\catcode `\#12\catcode `\^12\catcode `\_12\catcode `\%12\relax}%
\providecommand \@@startlink[1]{}%
\providecommand \@@endlink[0]{}%
\providecommand \url  [0]{\begingroup\@sanitize@url \@url }%
\providecommand \@url [1]{\endgroup\@href {#1}{\urlprefix }}%
\providecommand \urlprefix  [0]{URL }%
\providecommand \Eprint [0]{\href }%
\providecommand \doibase [0]{http://dx.doi.org/}%
\providecommand \selectlanguage [0]{\@gobble}%
\providecommand \bibinfo  [0]{\@secondoftwo}%
\providecommand \bibfield  [0]{\@secondoftwo}%
\providecommand \translation [1]{[#1]}%
\providecommand \BibitemOpen [0]{}%
\providecommand \bibitemStop [0]{}%
\providecommand \bibitemNoStop [0]{.\EOS\space}%
\providecommand \EOS [0]{\spacefactor3000\relax}%
\providecommand \BibitemShut  [1]{\csname bibitem#1\endcsname}%
\let\auto@bib@innerbib\@empty
\bibitem [{\citenamefont {Abbott}\ \emph
  {et~al.}(2019{\natexlab{a}})\citenamefont {Abbott} \emph
  {et~al.}}]{LIGOScientific:2018mvr}%
  \BibitemOpen
  \bibfield  {author} {\bibinfo {author} {\bibfnamefont {B.~P.}\ \bibnamefont
  {Abbott}} \emph {et~al.} (\bibinfo {collaboration} {LIGO Scientific,
  Virgo}),\ }\href {\doibase 10.1103/PhysRevX.9.031040} {\bibfield  {journal}
  {\bibinfo  {journal} {Phys. Rev.}\ }\textbf {\bibinfo {volume} {X9}},\
  \bibinfo {pages} {031040} (\bibinfo {year} {2019}{\natexlab{a}})},\ \Eprint
  {http://arxiv.org/abs/1811.12907} {arXiv:1811.12907 [astro-ph.HE]}
  \BibitemShut {NoStop}%
\bibitem [{\citenamefont {Abbott}\ \emph
  {et~al.}(2019{\natexlab{b}})\citenamefont {Abbott} \emph
  {et~al.}}]{LIGOScientific:2019fpa}%
  \BibitemOpen
  \bibfield  {author} {\bibinfo {author} {\bibfnamefont {B.~P.}\ \bibnamefont
  {Abbott}} \emph {et~al.} (\bibinfo {collaboration} {LIGO Scientific,
  Virgo}),\ }\href {\doibase 10.1103/PhysRevD.100.104036} {\bibfield  {journal}
  {\bibinfo  {journal} {Phys. Rev.}\ }\textbf {\bibinfo {volume} {D100}},\
  \bibinfo {pages} {104036} (\bibinfo {year} {2019}{\natexlab{b}})},\ \Eprint
  {http://arxiv.org/abs/1903.04467} {arXiv:1903.04467 [gr-qc]} \BibitemShut
  {NoStop}%
\bibitem [{\citenamefont {Abbott}\ \emph
  {et~al.}(2016{\natexlab{a}})\citenamefont {Abbott} \emph
  {et~al.}}]{TheLIGOScientific:2016pea}%
  \BibitemOpen
  \bibfield  {author} {\bibinfo {author} {\bibfnamefont {B.~P.}\ \bibnamefont
  {Abbott}} \emph {et~al.} (\bibinfo {collaboration} {LIGO Scientific,
  Virgo}),\ }\href {\doibase 10.1103/PhysRevX.6.041015,
  10.1103/PhysRevX.8.039903} {\bibfield  {journal} {\bibinfo  {journal} {Phys.
  Rev.}\ }\textbf {\bibinfo {volume} {X6}},\ \bibinfo {pages} {041015}
  (\bibinfo {year} {2016}{\natexlab{a}})},\ \bibinfo {note} {[erratum: Phys.
  Rev.X8,no.3,039903(2018)]},\ \Eprint {http://arxiv.org/abs/1606.04856}
  {arXiv:1606.04856 [gr-qc]} \BibitemShut {NoStop}%
\bibitem [{\citenamefont {Abbott}\ \emph
  {et~al.}(2016{\natexlab{b}})\citenamefont {Abbott} \emph
  {et~al.}}]{TheLIGOScientific:2016src}%
  \BibitemOpen
  \bibfield  {author} {\bibinfo {author} {\bibfnamefont {B.~P.}\ \bibnamefont
  {Abbott}} \emph {et~al.} (\bibinfo {collaboration} {LIGO Scientific,
  Virgo}),\ }\href {\doibase 10.1103/PhysRevLett.116.221101,
  10.1103/PhysRevLett.121.129902} {\bibfield  {journal} {\bibinfo  {journal}
  {Phys. Rev. Lett.}\ }\textbf {\bibinfo {volume} {116}},\ \bibinfo {pages}
  {221101} (\bibinfo {year} {2016}{\natexlab{b}})},\ \bibinfo {note} {[Erratum:
  Phys. Rev. Lett.121,no.12,129902(2018)]},\ \Eprint
  {http://arxiv.org/abs/1602.03841} {arXiv:1602.03841 [gr-qc]} \BibitemShut
  {NoStop}%
\bibitem [{\citenamefont {Abbott}\ \emph {et~al.}(2017)\citenamefont {Abbott}
  \emph {et~al.}}]{Abbott:2017vtc}%
  \BibitemOpen
  \bibfield  {author} {\bibinfo {author} {\bibfnamefont {B.~P.}\ \bibnamefont
  {Abbott}} \emph {et~al.} (\bibinfo {collaboration} {LIGO Scientific,
  VIRGO}),\ }\href {\doibase 10.1103/PhysRevLett.118.221101,
  10.1103/PhysRevLett.121.129901} {\bibfield  {journal} {\bibinfo  {journal}
  {Phys. Rev. Lett.}\ }\textbf {\bibinfo {volume} {118}},\ \bibinfo {pages}
  {221101} (\bibinfo {year} {2017})},\ \bibinfo {note} {[Erratum: Phys. Rev.
  Lett.121,no.12,129901(2018)]},\ \Eprint {http://arxiv.org/abs/1706.01812}
  {arXiv:1706.01812 [gr-qc]} \BibitemShut {NoStop}%
\bibitem [{\citenamefont {Lunin}\ and\ \citenamefont
  {Mathur}(2002)}]{Lunin:2001jy}%
  \BibitemOpen
  \bibfield  {author} {\bibinfo {author} {\bibfnamefont {O.}~\bibnamefont
  {Lunin}}\ and\ \bibinfo {author} {\bibfnamefont {S.~D.}\ \bibnamefont
  {Mathur}},\ }\href {\doibase 10.1016/S0550-3213(01)00620-4} {\bibfield
  {journal} {\bibinfo  {journal} {Nucl. Phys.}\ }\textbf {\bibinfo {volume}
  {B623}},\ \bibinfo {pages} {342} (\bibinfo {year} {2002})},\ \Eprint
  {http://arxiv.org/abs/hep-th/0109154} {arXiv:hep-th/0109154 [hep-th]}
  \BibitemShut {NoStop}%
\bibitem [{\citenamefont {Almheiri}\ \emph {et~al.}(2013)\citenamefont
  {Almheiri}, \citenamefont {Marolf}, \citenamefont {Polchinski},\ and\
  \citenamefont {Sully}}]{Almheiri:2012rt}%
  \BibitemOpen
  \bibfield  {author} {\bibinfo {author} {\bibfnamefont {A.}~\bibnamefont
  {Almheiri}}, \bibinfo {author} {\bibfnamefont {D.}~\bibnamefont {Marolf}},
  \bibinfo {author} {\bibfnamefont {J.}~\bibnamefont {Polchinski}}, \ and\
  \bibinfo {author} {\bibfnamefont {J.}~\bibnamefont {Sully}},\ }\href
  {\doibase 10.1007/JHEP02(2013)062} {\bibfield  {journal} {\bibinfo  {journal}
  {JHEP}\ }\textbf {\bibinfo {volume} {02}},\ \bibinfo {pages} {062} (\bibinfo
  {year} {2013})},\ \Eprint {http://arxiv.org/abs/1207.3123} {arXiv:1207.3123
  [hep-th]} \BibitemShut {NoStop}%
\bibitem [{\citenamefont {Mazur}\ and\ \citenamefont
  {Mottola}(2004)}]{Mazur:2004fk}%
  \BibitemOpen
  \bibfield  {author} {\bibinfo {author} {\bibfnamefont {P.~O.}\ \bibnamefont
  {Mazur}}\ and\ \bibinfo {author} {\bibfnamefont {E.}~\bibnamefont
  {Mottola}},\ }\href {\doibase 10.1073/pnas.0402717101} {\bibfield  {journal}
  {\bibinfo  {journal} {Proc. Nat. Acad. Sci.}\ }\textbf {\bibinfo {volume}
  {101}},\ \bibinfo {pages} {9545} (\bibinfo {year} {2004})},\ \Eprint
  {http://arxiv.org/abs/gr-qc/0407075} {arXiv:gr-qc/0407075 [gr-qc]}
  \BibitemShut {NoStop}%
\bibitem [{\citenamefont {Liebling}\ and\ \citenamefont
  {Palenzuela}(2012)}]{Liebling:2012fv}%
  \BibitemOpen
  \bibfield  {author} {\bibinfo {author} {\bibfnamefont {S.~L.}\ \bibnamefont
  {Liebling}}\ and\ \bibinfo {author} {\bibfnamefont {C.}~\bibnamefont
  {Palenzuela}},\ }\href {\doibase 10.12942/lrr-2012-6,
  10.1007/s41114-017-0007-y} {\bibfield  {journal} {\bibinfo  {journal} {Living
  Rev. Rel.}\ }\textbf {\bibinfo {volume} {15}},\ \bibinfo {pages} {6}
  (\bibinfo {year} {2012})},\ \bibinfo {note} {[Living Rev.
  Rel.20,no.1,5(2017)]},\ \Eprint {http://arxiv.org/abs/1202.5809}
  {arXiv:1202.5809 [gr-qc]} \BibitemShut {NoStop}%
\bibitem [{\citenamefont {Cardoso}\ \emph
  {et~al.}(2016{\natexlab{a}})\citenamefont {Cardoso}, \citenamefont
  {Franzin},\ and\ \citenamefont {Pani}}]{Cardoso:2016rao}%
  \BibitemOpen
  \bibfield  {author} {\bibinfo {author} {\bibfnamefont {V.}~\bibnamefont
  {Cardoso}}, \bibinfo {author} {\bibfnamefont {E.}~\bibnamefont {Franzin}}, \
  and\ \bibinfo {author} {\bibfnamefont {P.}~\bibnamefont {Pani}},\ }\href
  {\doibase 10.1103/PhysRevLett.116.171101} {\bibfield  {journal} {\bibinfo
  {journal} {Phys. Rev. Lett.}\ }\textbf {\bibinfo {volume} {116}},\ \bibinfo
  {pages} {171101} (\bibinfo {year} {2016}{\natexlab{a}})},\ \bibinfo {note}
  {[Erratum: Phys.Rev.Lett. 117, 089902 (2016)]},\ \Eprint
  {http://arxiv.org/abs/1602.07309} {arXiv:1602.07309 [gr-qc]} \BibitemShut
  {NoStop}%
\bibitem [{\citenamefont {Cardoso}\ \emph
  {et~al.}(2016{\natexlab{b}})\citenamefont {Cardoso}, \citenamefont {Hopper},
  \citenamefont {Macedo}, \citenamefont {Palenzuela},\ and\ \citenamefont
  {Pani}}]{Cardoso:2016oxy}%
  \BibitemOpen
  \bibfield  {author} {\bibinfo {author} {\bibfnamefont {V.}~\bibnamefont
  {Cardoso}}, \bibinfo {author} {\bibfnamefont {S.}~\bibnamefont {Hopper}},
  \bibinfo {author} {\bibfnamefont {C.~F.~B.}\ \bibnamefont {Macedo}}, \bibinfo
  {author} {\bibfnamefont {C.}~\bibnamefont {Palenzuela}}, \ and\ \bibinfo
  {author} {\bibfnamefont {P.}~\bibnamefont {Pani}},\ }\href {\doibase
  10.1103/PhysRevD.94.084031} {\bibfield  {journal} {\bibinfo  {journal} {Phys.
  Rev.}\ }\textbf {\bibinfo {volume} {D94}},\ \bibinfo {pages} {084031}
  (\bibinfo {year} {2016}{\natexlab{b}})},\ \Eprint
  {http://arxiv.org/abs/1608.08637} {arXiv:1608.08637 [gr-qc]} \BibitemShut
  {NoStop}%
\bibitem [{\citenamefont {Maselli}\ \emph {et~al.}(2018)\citenamefont
  {Maselli}, \citenamefont {Pani}, \citenamefont {Cardoso}, \citenamefont
  {Abdelsalhin}, \citenamefont {Gualtieri},\ and\ \citenamefont
  {Ferrari}}]{Maselli:2017cmm}%
  \BibitemOpen
  \bibfield  {author} {\bibinfo {author} {\bibfnamefont {A.}~\bibnamefont
  {Maselli}}, \bibinfo {author} {\bibfnamefont {P.}~\bibnamefont {Pani}},
  \bibinfo {author} {\bibfnamefont {V.}~\bibnamefont {Cardoso}}, \bibinfo
  {author} {\bibfnamefont {T.}~\bibnamefont {Abdelsalhin}}, \bibinfo {author}
  {\bibfnamefont {L.}~\bibnamefont {Gualtieri}}, \ and\ \bibinfo {author}
  {\bibfnamefont {V.}~\bibnamefont {Ferrari}},\ }\href {\doibase
  10.1103/PhysRevLett.120.081101} {\bibfield  {journal} {\bibinfo  {journal}
  {Phys. Rev. Lett.}\ }\textbf {\bibinfo {volume} {120}},\ \bibinfo {pages}
  {081101} (\bibinfo {year} {2018})},\ \Eprint
  {http://arxiv.org/abs/1703.10612} {arXiv:1703.10612 [gr-qc]} \BibitemShut
  {NoStop}%
\bibitem [{\citenamefont {Tsang}\ \emph {et~al.}(2020)\citenamefont {Tsang},
  \citenamefont {Ghosh}, \citenamefont {Samajdar}, \citenamefont
  {Chatziioannou}, \citenamefont {Mastrogiovanni}, \citenamefont {Agathos},\
  and\ \citenamefont {Van Den~Broeck}}]{Tsang:2019zra}%
  \BibitemOpen
  \bibfield  {author} {\bibinfo {author} {\bibfnamefont {K.~W.}\ \bibnamefont
  {Tsang}}, \bibinfo {author} {\bibfnamefont {A.}~\bibnamefont {Ghosh}},
  \bibinfo {author} {\bibfnamefont {A.}~\bibnamefont {Samajdar}}, \bibinfo
  {author} {\bibfnamefont {K.}~\bibnamefont {Chatziioannou}}, \bibinfo {author}
  {\bibfnamefont {S.}~\bibnamefont {Mastrogiovanni}}, \bibinfo {author}
  {\bibfnamefont {M.}~\bibnamefont {Agathos}}, \ and\ \bibinfo {author}
  {\bibfnamefont {C.}~\bibnamefont {Van Den~Broeck}},\ }\href {\doibase
  10.1103/PhysRevD.101.064012} {\bibfield  {journal} {\bibinfo  {journal}
  {Phys.\ Rev.\ D}\ }\textbf {\bibinfo {volume} {101}},\ \bibinfo {pages}
  {064012} (\bibinfo {year} {2020})},\ \Eprint
  {http://arxiv.org/abs/1906.11168} {arXiv:1906.11168 [gr-qc]} \BibitemShut
  {NoStop}%
\bibitem [{\citenamefont {Abedi}\ \emph {et~al.}(2017)\citenamefont {Abedi},
  \citenamefont {Dykaar},\ and\ \citenamefont {Afshordi}}]{Abedi:2016hgu}%
  \BibitemOpen
  \bibfield  {author} {\bibinfo {author} {\bibfnamefont {J.}~\bibnamefont
  {Abedi}}, \bibinfo {author} {\bibfnamefont {H.}~\bibnamefont {Dykaar}}, \
  and\ \bibinfo {author} {\bibfnamefont {N.}~\bibnamefont {Afshordi}},\ }\href
  {\doibase 10.1103/PhysRevD.96.082004} {\bibfield  {journal} {\bibinfo
  {journal} {Phys. Rev.}\ }\textbf {\bibinfo {volume} {D96}},\ \bibinfo {pages}
  {082004} (\bibinfo {year} {2017})},\ \Eprint
  {http://arxiv.org/abs/1612.00266} {arXiv:1612.00266 [gr-qc]} \BibitemShut
  {NoStop}%
\bibitem [{\citenamefont {Westerweck}\ \emph {et~al.}(2018)\citenamefont
  {Westerweck}, \citenamefont {Nielsen}, \citenamefont {Fischer-Birnholtz},
  \citenamefont {Cabero}, \citenamefont {Capano}, \citenamefont {Dent},
  \citenamefont {Krishnan}, \citenamefont {Meadors},\ and\ \citenamefont
  {Nitz}}]{Westerweck:2017hus}%
  \BibitemOpen
  \bibfield  {author} {\bibinfo {author} {\bibfnamefont {J.}~\bibnamefont
  {Westerweck}}, \bibinfo {author} {\bibfnamefont {A.}~\bibnamefont {Nielsen}},
  \bibinfo {author} {\bibfnamefont {O.}~\bibnamefont {Fischer-Birnholtz}},
  \bibinfo {author} {\bibfnamefont {M.}~\bibnamefont {Cabero}}, \bibinfo
  {author} {\bibfnamefont {C.}~\bibnamefont {Capano}}, \bibinfo {author}
  {\bibfnamefont {T.}~\bibnamefont {Dent}}, \bibinfo {author} {\bibfnamefont
  {B.}~\bibnamefont {Krishnan}}, \bibinfo {author} {\bibfnamefont
  {G.}~\bibnamefont {Meadors}}, \ and\ \bibinfo {author} {\bibfnamefont
  {A.~H.}\ \bibnamefont {Nitz}},\ }\href {\doibase 10.1103/PhysRevD.97.124037}
  {\bibfield  {journal} {\bibinfo  {journal} {Phys. Rev.}\ }\textbf {\bibinfo
  {volume} {D97}},\ \bibinfo {pages} {124037} (\bibinfo {year} {2018})},\
  \Eprint {http://arxiv.org/abs/1712.09966} {arXiv:1712.09966 [gr-qc]}
  \BibitemShut {NoStop}%
\bibitem [{\citenamefont {Cardoso}\ and\ \citenamefont
  {Pani}(2019)}]{Cardoso:2019rvt}%
  \BibitemOpen
  \bibfield  {author} {\bibinfo {author} {\bibfnamefont {V.}~\bibnamefont
  {Cardoso}}\ and\ \bibinfo {author} {\bibfnamefont {P.}~\bibnamefont {Pani}},\
  }\href@noop {} {\  (\bibinfo {year} {2019})},\ \Eprint
  {http://arxiv.org/abs/1904.05363} {arXiv:1904.05363 [gr-qc]} \BibitemShut
  {NoStop}%
\bibitem [{\citenamefont {Cardoso}\ \emph {et~al.}(2017)\citenamefont
  {Cardoso}, \citenamefont {Franzin}, \citenamefont {Maselli}, \citenamefont
  {Pani},\ and\ \citenamefont {Raposo}}]{Cardoso:2017cfl}%
  \BibitemOpen
  \bibfield  {author} {\bibinfo {author} {\bibfnamefont {V.}~\bibnamefont
  {Cardoso}}, \bibinfo {author} {\bibfnamefont {E.}~\bibnamefont {Franzin}},
  \bibinfo {author} {\bibfnamefont {A.}~\bibnamefont {Maselli}}, \bibinfo
  {author} {\bibfnamefont {P.}~\bibnamefont {Pani}}, \ and\ \bibinfo {author}
  {\bibfnamefont {G.}~\bibnamefont {Raposo}},\ }\href {\doibase
  10.1103/PhysRevD.95.089901, 10.1103/PhysRevD.95.084014} {\bibfield  {journal}
  {\bibinfo  {journal} {Phys. Rev.}\ }\textbf {\bibinfo {volume} {D95}},\
  \bibinfo {pages} {084014} (\bibinfo {year} {2017})},\ \bibinfo {note}
  {[Addendum: Phys. Rev.D95,no.8,089901(2017)]},\ \Eprint
  {http://arxiv.org/abs/1701.01116} {arXiv:1701.01116 [gr-qc]} \BibitemShut
  {NoStop}%
\bibitem [{\citenamefont {Sennett}\ \emph {et~al.}(2017)\citenamefont
  {Sennett}, \citenamefont {Hinderer}, \citenamefont {Steinhoff}, \citenamefont
  {Buonanno},\ and\ \citenamefont {Ossokine}}]{Sennett:2017etc}%
  \BibitemOpen
  \bibfield  {author} {\bibinfo {author} {\bibfnamefont {N.}~\bibnamefont
  {Sennett}}, \bibinfo {author} {\bibfnamefont {T.}~\bibnamefont {Hinderer}},
  \bibinfo {author} {\bibfnamefont {J.}~\bibnamefont {Steinhoff}}, \bibinfo
  {author} {\bibfnamefont {A.}~\bibnamefont {Buonanno}}, \ and\ \bibinfo
  {author} {\bibfnamefont {S.}~\bibnamefont {Ossokine}},\ }\href {\doibase
  10.1103/PhysRevD.96.024002} {\bibfield  {journal} {\bibinfo  {journal} {Phys.
  Rev.}\ }\textbf {\bibinfo {volume} {D96}},\ \bibinfo {pages} {024002}
  (\bibinfo {year} {2017})},\ \Eprint {http://arxiv.org/abs/1704.08651}
  {arXiv:1704.08651 [gr-qc]} \BibitemShut {NoStop}%
\bibitem [{\citenamefont {Brustein}\ and\ \citenamefont
  {Sherf}(2020)}]{Brustein:2020tpg}%
  \BibitemOpen
  \bibfield  {author} {\bibinfo {author} {\bibfnamefont {R.}~\bibnamefont
  {Brustein}}\ and\ \bibinfo {author} {\bibfnamefont {Y.}~\bibnamefont
  {Sherf}},\ }\href@noop {} {\  (\bibinfo {year} {2020})},\ \Eprint
  {http://arxiv.org/abs/2008.02738} {arXiv:2008.02738 [gr-qc]} \BibitemShut
  {NoStop}%
\bibitem [{\citenamefont {Brustein}\ and\ \citenamefont
  {Sherf}(2021)}]{Brustein:2021bnw}%
  \BibitemOpen
  \bibfield  {author} {\bibinfo {author} {\bibfnamefont {R.}~\bibnamefont
  {Brustein}}\ and\ \bibinfo {author} {\bibfnamefont {Y.}~\bibnamefont
  {Sherf}},\ }\href@noop {} {\  (\bibinfo {year} {2021})},\ \Eprint
  {http://arxiv.org/abs/2104.06013} {arXiv:2104.06013 [gr-qc]} \BibitemShut
  {NoStop}%
\bibitem [{\citenamefont {Krishnendu}\ \emph {et~al.}(2017)\citenamefont
  {Krishnendu}, \citenamefont {Arun},\ and\ \citenamefont
  {Mishra}}]{Krishnendu:2017shb}%
  \BibitemOpen
  \bibfield  {author} {\bibinfo {author} {\bibfnamefont {N.~V.}\ \bibnamefont
  {Krishnendu}}, \bibinfo {author} {\bibfnamefont {K.~G.}\ \bibnamefont
  {Arun}}, \ and\ \bibinfo {author} {\bibfnamefont {C.~K.}\ \bibnamefont
  {Mishra}},\ }\href {\doibase 10.1103/PhysRevLett.119.091101} {\bibfield
  {journal} {\bibinfo  {journal} {Phys. Rev. Lett.}\ }\textbf {\bibinfo
  {volume} {119}},\ \bibinfo {pages} {091101} (\bibinfo {year} {2017})},\
  \Eprint {http://arxiv.org/abs/1701.06318} {arXiv:1701.06318 [gr-qc]}
  \BibitemShut {NoStop}%
\bibitem [{\citenamefont {Datta}\ and\ \citenamefont
  {Bose}(2019)}]{Datta:2019euh}%
  \BibitemOpen
  \bibfield  {author} {\bibinfo {author} {\bibfnamefont {S.}~\bibnamefont
  {Datta}}\ and\ \bibinfo {author} {\bibfnamefont {S.}~\bibnamefont {Bose}},\
  }\href {\doibase 10.1103/PhysRevD.99.084001} {\bibfield  {journal} {\bibinfo
  {journal} {Phys. Rev.}\ }\textbf {\bibinfo {volume} {D99}},\ \bibinfo {pages}
  {084001} (\bibinfo {year} {2019})},\ \Eprint
  {http://arxiv.org/abs/1902.01723} {arXiv:1902.01723 [gr-qc]} \BibitemShut
  {NoStop}%
\bibitem [{\citenamefont {Datta}(2021)}]{Datta:2021hvm}%
  \BibitemOpen
  \bibfield  {author} {\bibinfo {author} {\bibfnamefont {S.}~\bibnamefont
  {Datta}},\ }\href@noop {} {\  (\bibinfo {year} {2021})},\ \Eprint
  {http://arxiv.org/abs/2107.07258} {arXiv:2107.07258 [gr-qc]} \BibitemShut
  {NoStop}%
\bibitem [{\citenamefont {Thorne}\ \emph {et~al.}(1986)\citenamefont {Thorne},
  \citenamefont {Price},\ and\ \citenamefont {Macdonald}}]{MembraneParadigm}%
  \BibitemOpen
  \bibfield  {author} {\bibinfo {author} {\bibfnamefont {K.~S.}\ \bibnamefont
  {Thorne}}, \bibinfo {author} {\bibfnamefont {R.}~\bibnamefont {Price}}, \
  and\ \bibinfo {author} {\bibfnamefont {D.}~\bibnamefont {Macdonald}},\
  }\href@noop {} {\emph {\bibinfo {title} {{Black holes: the membrane
  paradigm}}}},\ edited by\ \bibinfo {editor} {\bibfnamefont {K.~S.}\
  \bibnamefont {Thorne}}\ (\bibinfo  {publisher} {Yale University Press},\
  \bibinfo {year} {1986})\BibitemShut {NoStop}%
\bibitem [{\citenamefont {Damour}(1982)}]{Damour_viscous}%
  \BibitemOpen
  \bibfield  {author} {\bibinfo {author} {\bibfnamefont {T.}~\bibnamefont
  {Damour}},\ }in\ \href@noop {} {\emph {\bibinfo {booktitle} {{Proceedings of
  the Second Marcel Grossmann Meeting ofGeneral Relativity, edited by R.
  Ruffini, North Holland, Amsterdam, 1982 pp 587-608}}}}\ (\bibinfo {year}
  {1982})\BibitemShut {NoStop}%
\bibitem [{\citenamefont {Poisson}(2009)}]{Poisson:2009di}%
  \BibitemOpen
  \bibfield  {author} {\bibinfo {author} {\bibfnamefont {E.}~\bibnamefont
  {Poisson}},\ }\href {\doibase 10.1103/PhysRevD.80.064029} {\bibfield
  {journal} {\bibinfo  {journal} {Phys. Rev.}\ }\textbf {\bibinfo {volume}
  {D80}},\ \bibinfo {pages} {064029} (\bibinfo {year} {2009})},\ \Eprint
  {http://arxiv.org/abs/0907.0874} {arXiv:0907.0874 [gr-qc]} \BibitemShut
  {NoStop}%
\bibitem [{\citenamefont {Cardoso}\ and\ \citenamefont
  {Pani}(2013)}]{Cardoso:2012zn}%
  \BibitemOpen
  \bibfield  {author} {\bibinfo {author} {\bibfnamefont {V.}~\bibnamefont
  {Cardoso}}\ and\ \bibinfo {author} {\bibfnamefont {P.}~\bibnamefont {Pani}},\
  }\href {\doibase 10.1088/0264-9381/30/4/045011} {\bibfield  {journal}
  {\bibinfo  {journal} {Class. Quant. Grav.}\ }\textbf {\bibinfo {volume}
  {30}},\ \bibinfo {pages} {045011} (\bibinfo {year} {2013})},\ \Eprint
  {http://arxiv.org/abs/1205.3184} {arXiv:1205.3184 [gr-qc]} \BibitemShut
  {NoStop}%
\bibitem [{\citenamefont {Berti}\ \emph {et~al.}(2009)\citenamefont {Berti},
  \citenamefont {Cardoso},\ and\ \citenamefont {Starinets}}]{Berti:2009kk}%
  \BibitemOpen
  \bibfield  {author} {\bibinfo {author} {\bibfnamefont {E.}~\bibnamefont
  {Berti}}, \bibinfo {author} {\bibfnamefont {V.}~\bibnamefont {Cardoso}}, \
  and\ \bibinfo {author} {\bibfnamefont {A.~O.}\ \bibnamefont {Starinets}},\
  }\href {\doibase 10.1088/0264-9381/26/16/163001} {\bibfield  {journal}
  {\bibinfo  {journal} {Class. Quant. Grav.}\ }\textbf {\bibinfo {volume}
  {26}},\ \bibinfo {pages} {163001} (\bibinfo {year} {2009})},\ \Eprint
  {http://arxiv.org/abs/0905.2975} {arXiv:0905.2975 [gr-qc]} \BibitemShut
  {NoStop}%
\bibitem [{\citenamefont {Cardoso}\ \emph {et~al.}(2019)\citenamefont
  {Cardoso}, \citenamefont {Foit},\ and\ \citenamefont
  {Kleban}}]{Cardoso:2019apo}%
  \BibitemOpen
  \bibfield  {author} {\bibinfo {author} {\bibfnamefont {V.}~\bibnamefont
  {Cardoso}}, \bibinfo {author} {\bibfnamefont {V.~F.}\ \bibnamefont {Foit}}, \
  and\ \bibinfo {author} {\bibfnamefont {M.}~\bibnamefont {Kleban}},\ }\href
  {\doibase 10.1088/1475-7516/2019/08/006} {\bibfield  {journal} {\bibinfo
  {journal} {JCAP}\ }\textbf {\bibinfo {volume} {08}},\ \bibinfo {pages} {006}
  (\bibinfo {year} {2019})},\ \Eprint {http://arxiv.org/abs/1902.10164}
  {arXiv:1902.10164 [hep-th]} \BibitemShut {NoStop}%
\bibitem [{\citenamefont {Laghi}\ \emph {et~al.}(2021)\citenamefont {Laghi},
  \citenamefont {Carullo}, \citenamefont {Veitch},\ and\ \citenamefont
  {Del~Pozzo}}]{Laghi:2020rgl}%
  \BibitemOpen
  \bibfield  {author} {\bibinfo {author} {\bibfnamefont {D.}~\bibnamefont
  {Laghi}}, \bibinfo {author} {\bibfnamefont {G.}~\bibnamefont {Carullo}},
  \bibinfo {author} {\bibfnamefont {J.}~\bibnamefont {Veitch}}, \ and\ \bibinfo
  {author} {\bibfnamefont {W.}~\bibnamefont {Del~Pozzo}},\ }\href {\doibase
  10.1088/1361-6382/abde19} {\bibfield  {journal} {\bibinfo  {journal} {Class.
  Quant. Grav.}\ }\textbf {\bibinfo {volume} {38}},\ \bibinfo {pages} {095005}
  (\bibinfo {year} {2021})},\ \Eprint {http://arxiv.org/abs/2011.03816}
  {arXiv:2011.03816 [gr-qc]} \BibitemShut {NoStop}%
\bibitem [{\citenamefont {Agullo}\ \emph {et~al.}(2021)\citenamefont {Agullo},
  \citenamefont {Cardoso}, \citenamefont {Rio}, \citenamefont {Maggiore},\ and\
  \citenamefont {Pullin}}]{Agullo:2020hxe}%
  \BibitemOpen
  \bibfield  {author} {\bibinfo {author} {\bibfnamefont {I.}~\bibnamefont
  {Agullo}}, \bibinfo {author} {\bibfnamefont {V.}~\bibnamefont {Cardoso}},
  \bibinfo {author} {\bibfnamefont {A.~D.}\ \bibnamefont {Rio}}, \bibinfo
  {author} {\bibfnamefont {M.}~\bibnamefont {Maggiore}}, \ and\ \bibinfo
  {author} {\bibfnamefont {J.}~\bibnamefont {Pullin}},\ }\href {\doibase
  10.1103/PhysRevLett.126.041302} {\bibfield  {journal} {\bibinfo  {journal}
  {Phys. Rev. Lett.}\ }\textbf {\bibinfo {volume} {126}},\ \bibinfo {pages}
  {041302} (\bibinfo {year} {2021})},\ \Eprint
  {http://arxiv.org/abs/2007.13761} {arXiv:2007.13761 [gr-qc]} \BibitemShut
  {NoStop}%
\bibitem [{\citenamefont {Mukhanov}(1986)}]{Mukhanov:1986me}%
  \BibitemOpen
  \bibfield  {author} {\bibinfo {author} {\bibfnamefont {V.~F.}\ \bibnamefont
  {Mukhanov}},\ }\href@noop {} {\bibfield  {journal} {\bibinfo  {journal} {JETP
  Lett.}\ }\textbf {\bibinfo {volume} {44}},\ \bibinfo {pages} {63} (\bibinfo
  {year} {1986})}\BibitemShut {NoStop}%
\bibitem [{\citenamefont {Bekenstein}(1974)}]{Bekenstein:1974jk}%
  \BibitemOpen
  \bibfield  {author} {\bibinfo {author} {\bibfnamefont {J.~D.}\ \bibnamefont
  {Bekenstein}},\ }\href {\doibase 10.1007/BF02762768} {\bibfield  {journal}
  {\bibinfo  {journal} {Lett. Nuovo Cim.}\ }\textbf {\bibinfo {volume} {11}},\
  \bibinfo {pages} {467} (\bibinfo {year} {1974})}\BibitemShut {NoStop}%
\bibitem [{\citenamefont {Bekenstein}\ and\ \citenamefont
  {Mukhanov}(1995)}]{Bekenstein:1995ju}%
  \BibitemOpen
  \bibfield  {author} {\bibinfo {author} {\bibfnamefont {J.~D.}\ \bibnamefont
  {Bekenstein}}\ and\ \bibinfo {author} {\bibfnamefont {V.~F.}\ \bibnamefont
  {Mukhanov}},\ }\href {\doibase 10.1016/0370-2693(95)01148-J} {\bibfield
  {journal} {\bibinfo  {journal} {Phys. Lett. B}\ }\textbf {\bibinfo {volume}
  {360}},\ \bibinfo {pages} {7} (\bibinfo {year} {1995})},\ \Eprint
  {http://arxiv.org/abs/gr-qc/9505012} {arXiv:gr-qc/9505012} \BibitemShut
  {NoStop}%
\bibitem [{\citenamefont {Kothawala}\ \emph {et~al.}(2008)\citenamefont
  {Kothawala}, \citenamefont {Padmanabhan},\ and\ \citenamefont
  {Sarkar}}]{Kothawala:2008in}%
  \BibitemOpen
  \bibfield  {author} {\bibinfo {author} {\bibfnamefont {D.}~\bibnamefont
  {Kothawala}}, \bibinfo {author} {\bibfnamefont {T.}~\bibnamefont
  {Padmanabhan}}, \ and\ \bibinfo {author} {\bibfnamefont {S.}~\bibnamefont
  {Sarkar}},\ }\href {\doibase 10.1103/PhysRevD.78.104018} {\bibfield
  {journal} {\bibinfo  {journal} {Phys. Rev. D}\ }\textbf {\bibinfo {volume}
  {78}},\ \bibinfo {pages} {104018} (\bibinfo {year} {2008})},\ \Eprint
  {http://arxiv.org/abs/0807.1481} {arXiv:0807.1481 [gr-qc]} \BibitemShut
  {NoStop}%
\bibitem [{\citenamefont {Davidson}(2019)}]{Davidson:2019bqu}%
  \BibitemOpen
  \bibfield  {author} {\bibinfo {author} {\bibfnamefont {A.}~\bibnamefont
  {Davidson}},\ }\href {\doibase 10.1103/PhysRevD.100.081502} {\bibfield
  {journal} {\bibinfo  {journal} {Phys. Rev. D}\ }\textbf {\bibinfo {volume}
  {100}},\ \bibinfo {pages} {081502} (\bibinfo {year} {2019})},\ \Eprint
  {http://arxiv.org/abs/1907.03090} {arXiv:1907.03090 [gr-qc]} \BibitemShut
  {NoStop}%
\bibitem [{\citenamefont {de~Freitas~Pacheco}\ and\ \citenamefont
  {Silk}(2020)}]{deFreitasPacheco:2020wdg}%
  \BibitemOpen
  \bibfield  {author} {\bibinfo {author} {\bibfnamefont {J.~A.}\ \bibnamefont
  {de~Freitas~Pacheco}}\ and\ \bibinfo {author} {\bibfnamefont
  {J.}~\bibnamefont {Silk}},\ }\href {\doibase 10.1103/PhysRevD.101.083022}
  {\bibfield  {journal} {\bibinfo  {journal} {Phys. Rev. D}\ }\textbf {\bibinfo
  {volume} {101}},\ \bibinfo {pages} {083022} (\bibinfo {year} {2020})},\
  \Eprint {http://arxiv.org/abs/2003.12072} {arXiv:2003.12072 [astro-ph.CO]}
  \BibitemShut {NoStop}%
\bibitem [{\citenamefont {Agullo}\ \emph {et~al.}(2010)\citenamefont {Agullo},
  \citenamefont {Fernando~Barbero}, \citenamefont {Borja}, \citenamefont
  {Diaz-Polo},\ and\ \citenamefont {Villasenor}}]{Agullo:2010zz}%
  \BibitemOpen
  \bibfield  {author} {\bibinfo {author} {\bibfnamefont {I.}~\bibnamefont
  {Agullo}}, \bibinfo {author} {\bibfnamefont {J.}~\bibnamefont
  {Fernando~Barbero}}, \bibinfo {author} {\bibfnamefont {E.~F.}\ \bibnamefont
  {Borja}}, \bibinfo {author} {\bibfnamefont {J.}~\bibnamefont {Diaz-Polo}}, \
  and\ \bibinfo {author} {\bibfnamefont {E.~J.~S.}\ \bibnamefont
  {Villasenor}},\ }\href {\doibase 10.1103/PhysRevD.82.084029} {\bibfield
  {journal} {\bibinfo  {journal} {Phys. Rev. D}\ }\textbf {\bibinfo {volume}
  {82}},\ \bibinfo {pages} {084029} (\bibinfo {year} {2010})},\ \Eprint
  {http://arxiv.org/abs/1101.3660} {arXiv:1101.3660 [gr-qc]} \BibitemShut
  {NoStop}%
\bibitem [{\citenamefont {Rovelli}\ and\ \citenamefont
  {Smolin}(1995)}]{Rovelli:1994ge}%
  \BibitemOpen
  \bibfield  {author} {\bibinfo {author} {\bibfnamefont {C.}~\bibnamefont
  {Rovelli}}\ and\ \bibinfo {author} {\bibfnamefont {L.}~\bibnamefont
  {Smolin}},\ }\href {\doibase 10.1016/0550-3213(95)00150-Q} {\bibfield
  {journal} {\bibinfo  {journal} {Nucl. Phys. B}\ }\textbf {\bibinfo {volume}
  {442}},\ \bibinfo {pages} {593} (\bibinfo {year} {1995})},\ \bibinfo {note}
  {[Erratum: Nucl.Phys.B 456, 753--754 (1995)]},\ \Eprint
  {http://arxiv.org/abs/gr-qc/9411005} {arXiv:gr-qc/9411005} \BibitemShut
  {NoStop}%
\bibitem [{\citenamefont {Rovelli}(1996)}]{Rovelli:1996dv}%
  \BibitemOpen
  \bibfield  {author} {\bibinfo {author} {\bibfnamefont {C.}~\bibnamefont
  {Rovelli}},\ }\href {\doibase 10.1103/PhysRevLett.77.3288} {\bibfield
  {journal} {\bibinfo  {journal} {Phys. Rev. Lett.}\ }\textbf {\bibinfo
  {volume} {77}},\ \bibinfo {pages} {3288} (\bibinfo {year} {1996})},\ \Eprint
  {http://arxiv.org/abs/gr-qc/9603063} {arXiv:gr-qc/9603063} \BibitemShut
  {NoStop}%
\bibitem [{\citenamefont {Ashtekar}\ \emph {et~al.}(1998)\citenamefont
  {Ashtekar}, \citenamefont {Baez}, \citenamefont {Corichi},\ and\
  \citenamefont {Krasnov}}]{Ashtekar:1997yu}%
  \BibitemOpen
  \bibfield  {author} {\bibinfo {author} {\bibfnamefont {A.}~\bibnamefont
  {Ashtekar}}, \bibinfo {author} {\bibfnamefont {J.}~\bibnamefont {Baez}},
  \bibinfo {author} {\bibfnamefont {A.}~\bibnamefont {Corichi}}, \ and\
  \bibinfo {author} {\bibfnamefont {K.}~\bibnamefont {Krasnov}},\ }\href
  {\doibase 10.1103/PhysRevLett.80.904} {\bibfield  {journal} {\bibinfo
  {journal} {Phys. Rev. Lett.}\ }\textbf {\bibinfo {volume} {80}},\ \bibinfo
  {pages} {904} (\bibinfo {year} {1998})},\ \Eprint
  {http://arxiv.org/abs/gr-qc/9710007} {arXiv:gr-qc/9710007} \BibitemShut
  {NoStop}%
\bibitem [{\citenamefont {Ashtekar}\ \emph {et~al.}(2000)\citenamefont
  {Ashtekar}, \citenamefont {Baez},\ and\ \citenamefont
  {Krasnov}}]{Ashtekar:2000eq}%
  \BibitemOpen
  \bibfield  {author} {\bibinfo {author} {\bibfnamefont {A.}~\bibnamefont
  {Ashtekar}}, \bibinfo {author} {\bibfnamefont {J.~C.}\ \bibnamefont {Baez}},
  \ and\ \bibinfo {author} {\bibfnamefont {K.}~\bibnamefont {Krasnov}},\ }\href
  {\doibase 10.4310/ATMP.2000.v4.n1.a1} {\bibfield  {journal} {\bibinfo
  {journal} {Adv. Theor. Math. Phys.}\ }\textbf {\bibinfo {volume} {4}},\
  \bibinfo {pages} {1} (\bibinfo {year} {2000})},\ \Eprint
  {http://arxiv.org/abs/gr-qc/0005126} {arXiv:gr-qc/0005126} \BibitemShut
  {NoStop}%
\bibitem [{\citenamefont {Agullo}\ \emph {et~al.}(2008)\citenamefont {Agullo},
  \citenamefont {Barbero~G.}, \citenamefont {Diaz-Polo}, \citenamefont
  {Fernandez-Borja},\ and\ \citenamefont {Villasenor}}]{Agullo:2008yv}%
  \BibitemOpen
  \bibfield  {author} {\bibinfo {author} {\bibfnamefont {I.}~\bibnamefont
  {Agullo}}, \bibinfo {author} {\bibfnamefont {J.~F.}\ \bibnamefont
  {Barbero~G.}}, \bibinfo {author} {\bibfnamefont {J.}~\bibnamefont
  {Diaz-Polo}}, \bibinfo {author} {\bibfnamefont {E.}~\bibnamefont
  {Fernandez-Borja}}, \ and\ \bibinfo {author} {\bibfnamefont {E.~J.~S.}\
  \bibnamefont {Villasenor}},\ }\href {\doibase 10.1103/PhysRevLett.100.211301}
  {\bibfield  {journal} {\bibinfo  {journal} {Phys. Rev. Lett.}\ }\textbf
  {\bibinfo {volume} {100}},\ \bibinfo {pages} {211301} (\bibinfo {year}
  {2008})},\ \Eprint {http://arxiv.org/abs/0802.4077} {arXiv:0802.4077 [gr-qc]}
  \BibitemShut {NoStop}%
\bibitem [{\citenamefont {Chakraborty}\ and\ \citenamefont
  {Lochan}(2019)}]{Chakraborty:2017opo}%
  \BibitemOpen
  \bibfield  {author} {\bibinfo {author} {\bibfnamefont {S.}~\bibnamefont
  {Chakraborty}}\ and\ \bibinfo {author} {\bibfnamefont {K.}~\bibnamefont
  {Lochan}},\ }\href {\doibase 10.1016/j.physletb.2018.12.028} {\bibfield
  {journal} {\bibinfo  {journal} {Phys. Lett. B}\ }\textbf {\bibinfo {volume}
  {789}},\ \bibinfo {pages} {276} (\bibinfo {year} {2019})},\ \Eprint
  {http://arxiv.org/abs/1711.10660} {arXiv:1711.10660 [gr-qc]} \BibitemShut
  {NoStop}%
\bibitem [{\citenamefont {Brito}\ \emph {et~al.}(2015)\citenamefont {Brito},
  \citenamefont {Cardoso},\ and\ \citenamefont {Pani}}]{Brito:2015oca}%
  \BibitemOpen
  \bibfield  {author} {\bibinfo {author} {\bibfnamefont {R.}~\bibnamefont
  {Brito}}, \bibinfo {author} {\bibfnamefont {V.}~\bibnamefont {Cardoso}}, \
  and\ \bibinfo {author} {\bibfnamefont {P.}~\bibnamefont {Pani}},\ }\href
  {\doibase 10.1007/978-3-319-19000-6} {\bibfield  {journal} {\bibinfo
  {journal} {Lect. Notes Phys.}\ }\textbf {\bibinfo {volume} {906}},\ \bibinfo
  {pages} {pp.1} (\bibinfo {year} {2015})},\ \Eprint
  {http://arxiv.org/abs/1501.06570} {arXiv:1501.06570 [gr-qc]} \BibitemShut
  {NoStop}%
\bibitem [{\citenamefont {Hartle}(1973)}]{Hartle:1973zz}%
  \BibitemOpen
  \bibfield  {author} {\bibinfo {author} {\bibfnamefont {J.~B.}\ \bibnamefont
  {Hartle}},\ }\href {\doibase 10.1103/PhysRevD.8.1010} {\bibfield  {journal}
  {\bibinfo  {journal} {Phys. Rev.}\ }\textbf {\bibinfo {volume} {D8}},\
  \bibinfo {pages} {1010} (\bibinfo {year} {1973})}\BibitemShut {NoStop}%
\bibitem [{\citenamefont {Hughes}(2001)}]{Hughes:2001jr}%
  \BibitemOpen
  \bibfield  {author} {\bibinfo {author} {\bibfnamefont {S.~A.}\ \bibnamefont
  {Hughes}},\ }\href {\doibase 10.1103/PhysRevD.64.064004,
  10.1103/PhysRevD.88.109902} {\bibfield  {journal} {\bibinfo  {journal} {Phys.
  Rev.}\ }\textbf {\bibinfo {volume} {D64}},\ \bibinfo {pages} {064004}
  (\bibinfo {year} {2001})},\ \bibinfo {note} {[Erratum: Phys.
  Rev.D88,no.10,109902(2013)]},\ \Eprint {http://arxiv.org/abs/gr-qc/0104041}
  {arXiv:gr-qc/0104041 [gr-qc]} \BibitemShut {NoStop}%
\bibitem [{\citenamefont {Poisson}\ and\ \citenamefont
  {Will}(1953)}]{PoissonWill}%
  \BibitemOpen
  \bibfield  {author} {\bibinfo {author} {\bibfnamefont {E.}~\bibnamefont
  {Poisson}}\ and\ \bibinfo {author} {\bibfnamefont {C.}~\bibnamefont {Will}},\
  }\href@noop {} {\emph {\bibinfo {title} {{Gravity: Newtonian, Post-Newtonian,
  Relativistic}}}}\ (\bibinfo  {publisher} {Cambridge University Press},\
  \bibinfo {address} {Cambridge, UK},\ \bibinfo {year} {1953})\BibitemShut
  {NoStop}%
\bibitem [{\citenamefont {Datta}\ \emph
  {et~al.}(2020{\natexlab{a}})\citenamefont {Datta}, \citenamefont {Brito},
  \citenamefont {Bose}, \citenamefont {Pani},\ and\ \citenamefont
  {Hughes}}]{Datta:2019epe}%
  \BibitemOpen
  \bibfield  {author} {\bibinfo {author} {\bibfnamefont {S.}~\bibnamefont
  {Datta}}, \bibinfo {author} {\bibfnamefont {R.}~\bibnamefont {Brito}},
  \bibinfo {author} {\bibfnamefont {S.}~\bibnamefont {Bose}}, \bibinfo {author}
  {\bibfnamefont {P.}~\bibnamefont {Pani}}, \ and\ \bibinfo {author}
  {\bibfnamefont {S.~A.}\ \bibnamefont {Hughes}},\ }\href {\doibase
  10.1103/PhysRevD.101.044004} {\bibfield  {journal} {\bibinfo  {journal}
  {Phys. Rev.}\ }\textbf {\bibinfo {volume} {D101}},\ \bibinfo {pages} {044004}
  (\bibinfo {year} {2020}{\natexlab{a}})},\ \Eprint
  {http://arxiv.org/abs/1910.07841} {arXiv:1910.07841 [gr-qc]} \BibitemShut
  {NoStop}%
\bibitem [{\citenamefont {Datta}(2020)}]{Datta:2020rvo}%
  \BibitemOpen
  \bibfield  {author} {\bibinfo {author} {\bibfnamefont {S.}~\bibnamefont
  {Datta}},\ }\href {\doibase 10.1103/PhysRevD.102.064040} {\bibfield
  {journal} {\bibinfo  {journal} {Phys. Rev. D}\ }\textbf {\bibinfo {volume}
  {102}},\ \bibinfo {pages} {064040} (\bibinfo {year} {2020})},\ \Eprint
  {http://arxiv.org/abs/2002.04480} {arXiv:2002.04480 [gr-qc]} \BibitemShut
  {NoStop}%
\bibitem [{\citenamefont {Sherf}(2021)}]{Sherf:2021ppp}%
  \BibitemOpen
  \bibfield  {author} {\bibinfo {author} {\bibfnamefont {Y.}~\bibnamefont
  {Sherf}},\ }\href {\doibase 10.1103/PhysRevD.103.104003} {\bibfield
  {journal} {\bibinfo  {journal} {Phys. Rev. D}\ }\textbf {\bibinfo {volume}
  {103}},\ \bibinfo {pages} {104003} (\bibinfo {year} {2021})},\ \Eprint
  {http://arxiv.org/abs/2104.03766} {arXiv:2104.03766 [gr-qc]} \BibitemShut
  {NoStop}%
\bibitem [{\citenamefont {Chakraborty}\ \emph {et~al.}(2021)\citenamefont
  {Chakraborty}, \citenamefont {Datta},\ and\ \citenamefont
  {Sau}}]{Chakraborty:2021gdf}%
  \BibitemOpen
  \bibfield  {author} {\bibinfo {author} {\bibfnamefont {S.}~\bibnamefont
  {Chakraborty}}, \bibinfo {author} {\bibfnamefont {S.}~\bibnamefont {Datta}},
  \ and\ \bibinfo {author} {\bibfnamefont {S.}~\bibnamefont {Sau}},\
  }\href@noop {} {\  (\bibinfo {year} {2021})},\ \Eprint
  {http://arxiv.org/abs/2103.12430} {arXiv:2103.12430 [gr-qc]} \BibitemShut
  {NoStop}%
\bibitem [{\citenamefont {Maggio}\ \emph {et~al.}(2021)\citenamefont {Maggio},
  \citenamefont {van~de Meent},\ and\ \citenamefont {Pani}}]{Maggio:2021uge}%
  \BibitemOpen
  \bibfield  {author} {\bibinfo {author} {\bibfnamefont {E.}~\bibnamefont
  {Maggio}}, \bibinfo {author} {\bibfnamefont {M.}~\bibnamefont {van~de
  Meent}}, \ and\ \bibinfo {author} {\bibfnamefont {P.}~\bibnamefont {Pani}},\
  }\href@noop {} {\  (\bibinfo {year} {2021})},\ \Eprint
  {http://arxiv.org/abs/2106.07195} {arXiv:2106.07195 [gr-qc]} \BibitemShut
  {NoStop}%
\bibitem [{\citenamefont {Sago}\ and\ \citenamefont
  {Tanaka}(2021)}]{Sago:2021iku}%
  \BibitemOpen
  \bibfield  {author} {\bibinfo {author} {\bibfnamefont {N.}~\bibnamefont
  {Sago}}\ and\ \bibinfo {author} {\bibfnamefont {T.}~\bibnamefont {Tanaka}},\
  }\href@noop {} {\  (\bibinfo {year} {2021})},\ \Eprint
  {http://arxiv.org/abs/2106.07123} {arXiv:2106.07123 [gr-qc]} \BibitemShut
  {NoStop}%
\bibitem [{\citenamefont {Alvi}(2001)}]{Alvi:2001mx}%
  \BibitemOpen
  \bibfield  {author} {\bibinfo {author} {\bibfnamefont {K.}~\bibnamefont
  {Alvi}},\ }\href {\doibase 10.1103/PhysRevD.64.104020} {\bibfield  {journal}
  {\bibinfo  {journal} {Phys. Rev.}\ }\textbf {\bibinfo {volume} {D64}},\
  \bibinfo {pages} {104020} (\bibinfo {year} {2001})},\ \Eprint
  {http://arxiv.org/abs/gr-qc/0107080} {arXiv:gr-qc/0107080 [gr-qc]}
  \BibitemShut {NoStop}%
\bibitem [{\citenamefont {Datta}\ \emph
  {et~al.}(2020{\natexlab{b}})\citenamefont {Datta}, \citenamefont {Phukon},\
  and\ \citenamefont {Bose}}]{Datta:2020gem}%
  \BibitemOpen
  \bibfield  {author} {\bibinfo {author} {\bibfnamefont {S.}~\bibnamefont
  {Datta}}, \bibinfo {author} {\bibfnamefont {K.~S.}\ \bibnamefont {Phukon}}, \
  and\ \bibinfo {author} {\bibfnamefont {S.}~\bibnamefont {Bose}},\ }\href@noop
  {} {\  (\bibinfo {year} {2020}{\natexlab{b}})},\ \Eprint
  {http://arxiv.org/abs/2004.05974} {arXiv:2004.05974 [gr-qc]} \BibitemShut
  {NoStop}%
\bibitem [{\citenamefont {Chatziioannou}\ \emph {et~al.}(2013)\citenamefont
  {Chatziioannou}, \citenamefont {Poisson},\ and\ \citenamefont
  {Yunes}}]{Chatziioannou:2012gq}%
  \BibitemOpen
  \bibfield  {author} {\bibinfo {author} {\bibfnamefont {K.}~\bibnamefont
  {Chatziioannou}}, \bibinfo {author} {\bibfnamefont {E.}~\bibnamefont
  {Poisson}}, \ and\ \bibinfo {author} {\bibfnamefont {N.}~\bibnamefont
  {Yunes}},\ }\href {\doibase 10.1103/PhysRevD.87.044022} {\bibfield  {journal}
  {\bibinfo  {journal} {Phys. Rev.}\ }\textbf {\bibinfo {volume} {D87}},\
  \bibinfo {pages} {044022} (\bibinfo {year} {2013})},\ \Eprint
  {http://arxiv.org/abs/1211.1686} {arXiv:1211.1686 [gr-qc]} \BibitemShut
  {NoStop}%
\bibitem [{\citenamefont {Chatziioannou}\ \emph {et~al.}(2016)\citenamefont
  {Chatziioannou}, \citenamefont {Poisson},\ and\ \citenamefont
  {Yunes}}]{Chatziioannou:2016kem}%
  \BibitemOpen
  \bibfield  {author} {\bibinfo {author} {\bibfnamefont {K.}~\bibnamefont
  {Chatziioannou}}, \bibinfo {author} {\bibfnamefont {E.}~\bibnamefont
  {Poisson}}, \ and\ \bibinfo {author} {\bibfnamefont {N.}~\bibnamefont
  {Yunes}},\ }\href {\doibase 10.1103/PhysRevD.94.084043} {\bibfield  {journal}
  {\bibinfo  {journal} {Phys. Rev. D}\ }\textbf {\bibinfo {volume} {94}},\
  \bibinfo {pages} {084043} (\bibinfo {year} {2016})},\ \Eprint
  {http://arxiv.org/abs/1608.02899} {arXiv:1608.02899 [gr-qc]} \BibitemShut
  {NoStop}%
\bibitem [{\citenamefont {Tichy}\ \emph {et~al.}(2000)\citenamefont {Tichy},
  \citenamefont {Flanagan},\ and\ \citenamefont {Poisson}}]{Tichy:1999pv}%
  \BibitemOpen
  \bibfield  {author} {\bibinfo {author} {\bibfnamefont {W.}~\bibnamefont
  {Tichy}}, \bibinfo {author} {\bibfnamefont {E.~E.}\ \bibnamefont {Flanagan}},
  \ and\ \bibinfo {author} {\bibfnamefont {E.}~\bibnamefont {Poisson}},\ }\href
  {\doibase 10.1103/PhysRevD.61.104015} {\bibfield  {journal} {\bibinfo
  {journal} {Phys. Rev.}\ }\textbf {\bibinfo {volume} {D61}},\ \bibinfo {pages}
  {104015} (\bibinfo {year} {2000})},\ \Eprint
  {http://arxiv.org/abs/gr-qc/9912075} {arXiv:gr-qc/9912075 [gr-qc]}
  \BibitemShut {NoStop}%
\bibitem [{\citenamefont {Isoyama}\ and\ \citenamefont
  {Nakano}(2018)}]{Isoyama:2017tbp}%
  \BibitemOpen
  \bibfield  {author} {\bibinfo {author} {\bibfnamefont {S.}~\bibnamefont
  {Isoyama}}\ and\ \bibinfo {author} {\bibfnamefont {H.}~\bibnamefont
  {Nakano}},\ }\href {\doibase 10.1088/1361-6382/aa96c5} {\bibfield  {journal}
  {\bibinfo  {journal} {Class. Quant. Grav.}\ }\textbf {\bibinfo {volume}
  {35}},\ \bibinfo {pages} {024001} (\bibinfo {year} {2018})},\ \Eprint
  {http://arxiv.org/abs/1705.03869} {arXiv:1705.03869 [gr-qc]} \BibitemShut
  {NoStop}%
\bibitem [{\citenamefont {Taracchini}\ \emph {et~al.}(2013)\citenamefont
  {Taracchini}, \citenamefont {Buonanno}, \citenamefont {Hughes},\ and\
  \citenamefont {Khanna}}]{Taracchini:2013wfa}%
  \BibitemOpen
  \bibfield  {author} {\bibinfo {author} {\bibfnamefont {A.}~\bibnamefont
  {Taracchini}}, \bibinfo {author} {\bibfnamefont {A.}~\bibnamefont
  {Buonanno}}, \bibinfo {author} {\bibfnamefont {S.~A.}\ \bibnamefont
  {Hughes}}, \ and\ \bibinfo {author} {\bibfnamefont {G.}~\bibnamefont
  {Khanna}},\ }\href {\doibase 10.1103/PhysRevD.88.044001} {\bibfield
  {journal} {\bibinfo  {journal} {Phys. Rev. D}\ }\textbf {\bibinfo {volume}
  {88}},\ \bibinfo {pages} {044001} (\bibinfo {year} {2013})},\ \bibinfo {note}
  {[Erratum: Phys.Rev.D 88, 109903 (2013)]},\ \Eprint
  {http://arxiv.org/abs/1305.2184} {arXiv:1305.2184 [gr-qc]} \BibitemShut
  {NoStop}%
\bibitem [{\citenamefont {Chung}\ and\ \citenamefont
  {Sakellariadou}(2021)}]{Chung:2020uqj}%
  \BibitemOpen
  \bibfield  {author} {\bibinfo {author} {\bibfnamefont {A.~K.-W.}\
  \bibnamefont {Chung}}\ and\ \bibinfo {author} {\bibfnamefont
  {M.}~\bibnamefont {Sakellariadou}},\ }\href {\doibase
  10.1140/epjc/s10052-021-09391-3} {\bibfield  {journal} {\bibinfo  {journal}
  {Eur. Phys. J. C}\ }\textbf {\bibinfo {volume} {81}},\ \bibinfo {pages} {592}
  (\bibinfo {year} {2021})},\ \Eprint {http://arxiv.org/abs/2003.09778}
  {arXiv:2003.09778 [gr-qc]} \BibitemShut {NoStop}%
\bibitem [{\citenamefont {Maggio}\ \emph {et~al.}(2017)\citenamefont {Maggio},
  \citenamefont {Pani},\ and\ \citenamefont {Ferrari}}]{Maggio:2017ivp}%
  \BibitemOpen
  \bibfield  {author} {\bibinfo {author} {\bibfnamefont {E.}~\bibnamefont
  {Maggio}}, \bibinfo {author} {\bibfnamefont {P.}~\bibnamefont {Pani}}, \ and\
  \bibinfo {author} {\bibfnamefont {V.}~\bibnamefont {Ferrari}},\ }\href
  {\doibase 10.1103/PhysRevD.96.104047} {\bibfield  {journal} {\bibinfo
  {journal} {Phys. Rev. D}\ }\textbf {\bibinfo {volume} {96}},\ \bibinfo
  {pages} {104047} (\bibinfo {year} {2017})},\ \Eprint
  {http://arxiv.org/abs/1703.03696} {arXiv:1703.03696 [gr-qc]} \BibitemShut
  {NoStop}%
\bibitem [{\citenamefont {Maggio}\ \emph {et~al.}(2019)\citenamefont {Maggio},
  \citenamefont {Cardoso}, \citenamefont {Dolan},\ and\ \citenamefont
  {Pani}}]{Maggio:2018ivz}%
  \BibitemOpen
  \bibfield  {author} {\bibinfo {author} {\bibfnamefont {E.}~\bibnamefont
  {Maggio}}, \bibinfo {author} {\bibfnamefont {V.}~\bibnamefont {Cardoso}},
  \bibinfo {author} {\bibfnamefont {S.~R.}\ \bibnamefont {Dolan}}, \ and\
  \bibinfo {author} {\bibfnamefont {P.}~\bibnamefont {Pani}},\ }\href {\doibase
  10.1103/PhysRevD.99.064007} {\bibfield  {journal} {\bibinfo  {journal} {Phys.
  Rev. D}\ }\textbf {\bibinfo {volume} {99}},\ \bibinfo {pages} {064007}
  (\bibinfo {year} {2019})},\ \Eprint {http://arxiv.org/abs/1807.08840}
  {arXiv:1807.08840 [gr-qc]} \BibitemShut {NoStop}%
\bibitem [{\citenamefont {Chakravarti}\ \emph {et~al.}(2021)\citenamefont
  {Chakravarti}, \citenamefont {Ghosh},\ and\ \citenamefont
  {Sarkar}}]{Chakravarti:2021jbv}%
  \BibitemOpen
  \bibfield  {author} {\bibinfo {author} {\bibfnamefont {K.}~\bibnamefont
  {Chakravarti}}, \bibinfo {author} {\bibfnamefont {R.}~\bibnamefont {Ghosh}},
  \ and\ \bibinfo {author} {\bibfnamefont {S.}~\bibnamefont {Sarkar}},\
  }\href@noop {} {\  (\bibinfo {year} {2021})},\ \Eprint
  {http://arxiv.org/abs/2108.02444} {arXiv:2108.02444 [gr-qc]} \BibitemShut
  {NoStop}%
\end{thebibliography}%
\end{document}